\documentstyle[12pt]{article}
\newcommand{\bg}[1]{\mbox{\boldmath ${#1}$}}

\begin{document}
\begin{titlepage}
\thispagestyle{empty}

\vspace*{3mm}

\begin{flushright}
\large
{\sf to appear in} {\it Superconductivity Review,} {\sf 1995}
\end{flushright}

\vspace{15mm}

\begin{center}
{\LARGE
MAGNETIC PROPERTIES OF \\[2mm]
UNCONVENTIONAL SUPERCONDUCTORS}\\[11mm]
{\Large I.\,A.\,Luk'yanchuk and M.\,E.\,Zhitomirsky}  \\[3mm]
L.\,D.\,Landau Institute for Theoretical Physics, Moscow, 117334,
Russia \\[7mm]
5 August, 1994
\end{center}

\vspace{11mm}

\begin{abstract}
The article reviews recent developments on magnetic
properties of superconductors with anisotropic Cooper pairing. In
particular, we show how the concept of broken symmetries is applied to
the investigation of the mixed state in superconductors with a
multicomponent order parameter. Starting from the phenomenological
description in the framework of the generalized Ginzburg-Landau
theory, we discuss different types of quantized vortices appearing at
$H_{c1}$ in states with and without time-reversal breaking. General
classification of superconducting phase transitions in a uniform
magnetic field at $H_{c2}$ is constructed. Vortex lattices of
different forms are found in the vicinity of the upper critical field.
Symmetry arguments are used to classify phase transitions inside the
mixed state. Special attention is given to results which can be
obtained analytically. Also special emphasis is put on the open
questions of the theory.
\end{abstract}

\end{titlepage}

\section{Introduction}

     The BCS microscopic theory of superconductivity [$^{7}$] was
preceded by the phenomenological approach to the superconducting phase
transition developed in the works by Ginzburg, Landau [$^{20}$] and
Abrikosov [$^{1}$]. Considering the superconducting order parameter
as a scalar complex wave function which breaks gauge symmetry, one can
explain many remarkable properties of superconductors: the Meisner
effect, critical behavior of bulk samples and thin films, two types of
superconductivity, and so on. The microscopic approach developed later
clarified the physical meaning of the complex order parameter and
established the relation between phenomenological constants of the
Ginzburg-Landau (GL) functional and microscopic characteristics of
superconducting metal.

     One of the basic points of the BCS theory is the assumption that
electrons are paired in a fully isotropic {\it s-wave} state with both
total spin $S$ and orbital momentum $L$ of the Cooper pairs equal to
zero. The possibility of the {\it unconventional superconductivity}
with anisotropic, non $s$-wave pairing has been discussed since early
60th. In particular, it was shown [$^{10,57}$] that  anisotropic
Cooper pairs are formed if the interaction between fermion
quasiparticles is attractive for at least one value of the relative
orbital momentum.

     The feature of the phenomenological theory of unconventional
superconductors is that they are described by the {\it multicomponent}
order parameter with a more complicated structure of the GL
functional. This leads to qualitatively new properties of
unconventional superconductivity. The behavior of
superconductors with a multicomponent order parameter in magnetic
field is the subject of this review.

     Until the beginning of the 80th, superfluid $^3$He was the only
known example of an unconventional pairing. However, because of the
absence of electric charge, the interaction of superfluid
component in $^3$He with magnetic field is restricted to the weak spin
paramagnetism of Cooper pairs, and lies therefore beyond our
attention. (The complete review on superfluity of $^3$He is given in
[$^{73}$].)

     Since the discovery of superconductivity in so called heavy
fermion compounds [$^{17,23,51}$], they have been extensively
discussed as new possible candidates for anisotropic pairing. Some of
them, such as CeCu$_2$Si$_2$, UBe$_{13}$, URu$_2$Si$_2$, UPt$_3$,
UPd$_2$Al$_3$, UNi$_2$Al$_3$, reveal various unusual properties, which
could be attributed to unconventional superconducting states
[$^{36,66,68}$].

     The most convincing evidence of the nontrivial superconductivity
in heavy fermions is the complicated $H$--$T$ phase diagram of
superconducting UPt$_3$ (Fig.\ 1). In numerous experiments the
splitting of the superconducting phase transition in zero magnetic
field, the discontinuity in slope (kink) of $H_{c2}(T)$, two different
superconducting states at $H=0$ and three different vortex phases have
been observed (see [$^{36,68}$] and reference there in). Such
multiphase diagram could not be interpreted in terms of the usual
GL theory. It seems natural to explain it assuming
that superconductivity in UPt$_3$ is described by a
multicomponent order parameter with different minimums of
corresponding GL functional, and phase transitions take place between
them.

     A number of questions about magnetic properties of unconventional
superconductors are inspired by these experiments: what are the common
and different features of the unconventional and $s$-wave
superconductors, what are the phase transitions in the mixed state of
superconductors with a multicomponent order parameter? In this paper
we report on the recent progress in this field achieved since the
publication of previous reviews [$^{22,65}$]. We concentrate mostly
on theoretical aspects of the description of the vortex
state in unconventional superconductors (the list of the discussed
questions is given at the end of Sec.\ 1.3). We do not suppose to present
here details of the concrete models designed to explain peculiar
properties of UPt$_3$, this was done recently in a number of works
[$^{27,43,61}$]. However, we use these models as an
illustration of the general approach and try to stress, when it is possible,
which particular properties of the mixed state of multicomponent
superconductors can be used for reliable identification of an
unconventional superconductivity in heavy fermions.

     Three following Subsections have an introductorily character
and give the outline of the GL theory of unconventional
superconductors, the main points of the Abrikosov theory of the mixed
state of conventional $s$-wave superconductors and the properties of
multicomponent superconductors at $H=0$.

\subsection{Ginzburg-Landau Theory}

     In this Subsection we consider GL theory of unconventional
superconductivity developed in [$^{8,72,75}$]. We define the order
parameter and the symmetry group of the space uniform state of
unconventional superconductors (at $H=0$) which will be used to
describe the mixed state in the next Sections. The two-component
model, which is commonly used for the description of unconventional
superconductivity in UPt$_3$, is considered as a particular example of
unconventional superconductivity.

     The superconducting second order phase transition in the ordinary
case results in the breaking of normal state gauge symmetry. To
give the phenomenological description of the phase transition in an
unconventional superconductor, we first define the symmetry group of a
normal metal, which can be broken when the condensate of Cooper pairs
appears. This group includes:
\begin{itemize}
\item The gauge group $U(1)$ of multiplication of electron wave
      functions $\Psi$ by an arbitrary phase factor: $\Psi
      \rightarrow e^{i\alpha }\Psi$.
\item The space group of the crystal lattice.
\item The group of transformations of the spin space.
\item The time reversal operation $R$.
\end{itemize}
Two remarks are relevant at this point.

\noindent
(i) The space group, generally speaking, includes different
combinations of translations, rotations and reflections. However,
because of the large size of Cooper pairs (which at $T=0$ is of the
order of coherence length $\xi_0$) in comparison to lattice constant
$a$ (in BCS theory $a/\xi_0\sim T_c/E_F$), the discrete crystal
structure has no effect on the classification of superconducting
states. Therefore, a superconductor can be considered as uniform but
anisotropic medium with the space symmetry $T\times G$, where $T$ is
the abelian group of continuous three dimensional translations, and
$G$ is the point group of the crystal. Classification of superconducting
states were done taking into account discrete crystal structure in
[$^{52,47}$].

\noindent
(ii) Transformations of the spin space are separate from real space
transformations only if the spin-orbital interaction is negligible (as
in the case of superfluid $^3$He). In the opposite case of strong
spin-orbital coupling, one should assume that point group $G$ acts in
both coordinate and spin spaces. Since the strong spin-orbital
coupling seems to be relevant for properties of heavy fermion
compounds, we will concentrate on the latter case. Note, however, that
the spin-orbital interaction does not lift the Kramers degeneracy, and
superconducting electrons are characterized by the two-component wave
function $\Psi_\alpha$. This pseudospin degeneracy is important
for the triplet-singlet classification of the superconducting state
[$^{4,72,75}$] (see below).

Finally, the symmetry group of the normal state ${\cal G}$ can be
written as follows:
\begin{equation}
{\cal G}=T\times G\times R\times U(1)\ .
\label{G}
\end{equation}
The superconducting order parameter is the condensate wave function
\begin{equation}
\Delta_{\alpha \beta}({\bf r}_1,{\bf r}_2) =
\langle\hat{\Psi}_\alpha({\bf r}_1)\hat{\Psi}_\beta ({\bf r}_2)
\rangle\ .
\label{OP}
\end{equation}
It is more convenient to rewrite (\ref{OP}) in Wigner variables:
Cooper pair center of mass coordinate ${\bf r}=({\bf r}_1 + {\bf
r}_2)/2$ and position on the Fermi surface ${\bf k}$, which is
obtained by Fourier transformation: $({\bf r}_1-{\bf r}_2)\rightarrow
{\bf k}$. The electron energy spectrum is reconstructed in a thin
layer near the Fermi surface. Hence, the dependence of
$\Delta_{\alpha\beta}({\bf k},{\bf r})$ on the direction of $\bf k$ is
more important than on its modulus. In this Subsection we assume that
the superconducting state is homogeneous in space what means that
$\Delta_{\alpha \beta }$ does not depend on ${\bf r}$. For this reason
we will not discuss here translational properties of the order parameter.

     The possible superconducting states are classified according to
the symmetry of $\Delta_{\alpha\beta}({\bf k})$. We have already
mentioned that the superconducting phase transition breaks the gauge
invariance $U(1)$. Hence possible residual symmetries of
$\Delta_{\alpha\beta}({\bf k})$ may be found by listing all subgroups
of group $\cal G$ which do not contain the group $U(1)$ as a separate
subgroup [$^{8,75}$]. For a conventional superconductor this subgroup
is equal to $G\times R$. If it is smaller than $G\times R$, the order
parameter $\Delta_{\alpha \beta}({\bf k})$ is defined to be
unconventional.

     Not all subgroups of $\cal G$ correspond to energeticaly stable
superconducting phases. The application of the Landau symmetry
approach restricts the list of possible superconducting phases arising
just below $T_c$.

     First of all, the superconducting phase transition corresponds to
a particular irreducible representation of the point group $G$. More
exactly, the order parameter $\Delta_{\alpha\beta}({\bf k})$ should be
expanded over the basis functions $\Phi_{\alpha \beta }^i({\bf k})$ of
the given irreducible representation in the vicinity of the critical
temperature:
\begin{equation}
\Delta_{\alpha \beta}({\bf k},{\bf r}) = \sum_i\, \eta_i({\bf r})
\Phi_{\alpha \beta}^i({\bf k})\ .    \label{expan}
\end{equation}
Another important property of $\Delta_{\alpha\beta}({\bf k})$
follows from the anticommutation relation on fermion operators in
(\ref{OP}): $\Delta_{\alpha\beta}({\bf k}) = -
\Delta_{\beta\alpha}(-{\bf k})$. Therefore, for crystals with space inversion
symmetry, the superconducting pairing occurs either in the singlet,
space-even states with $\hat{\Delta}({\bf k}) = i
\hat{\sigma}_y\psi({\bf k})$ (where $\psi({\bf k})=\psi (-{\bf
k})$) or in the triplet, space-odd states with $\hat{\Delta} ({\bf
k}) = i \hat{\sigma }_y(\hat{\bg{\sigma}}\!\cdot \!{\bf d}({\bf
k}))$ (where ${\bf d}({\bf k})=-{\bf d}(-{\bf k})$). As the gap in the
quasiparticles energy spectrum is propotional to the superconducting
order parameter, the zeroes of the latter play important role for low
temperature thermodynamics of a superconductor [$^{22,65}$].
The symmetry analysis
[$^{8,75}$] shows that in the triplet case the types of zeros in the
energy gap always correspond to points on the Fermi surface, whereas
whole lines of zeros are possible for the singlet pairing.

     When this phenomenological approach is applied to heavy fermion
superconductors, the following one- (1D) and multidimensional (2D or
3D) irreducible representations for different point groups $G$ are
usually considered (we use notations of [$^{33}$]):
\smallskip

\noindent
1D: $A_1$, $A_2$, $B_1$, $B_2$ and 2D: $E_{1g}$, $E_{1u}$, $E_{2u}$
for the hexagonal group $D_{6h}$ of UPt$_3$, UPd$_2$Al$_3$,
UNi$_2$Al$_3$.

\noindent
1D: $A_1$, $A_2$, and 2D: $E$ for the tetragonal group
$D_{4h}$ of CeCu$_2$Si$_2$, URu$_2$Si$_2$.

\noindent
1D: $A_1$, $A_2$, 2D: $E$ and 3D: $F_1$, $F_2$ for the cubic group
$O_h$ of UBe$_{13}$.
\smallskip

\noindent
The explicit form of basis functions of these irreducible
representations is given in [$^{22,65}$].

     Near $T_c$ free energy can be obtained as an expansion in powers of
the order parameter which is invariant under the action of group $\cal G$.

     Consider first the case of 1D irreducible representation of $G$.
The symmetry of the order parameter is determined by the basis
function $\hat{\Phi}({\bf k})$. When this representation is identical,
$\hat{\Phi}({\bf k})$ has the full symmetry of the group $G\times R$
and corresponds to the conventional Cooper pairing. For nonidentical
representations corresponding to the unconventional pairing some
symmetry elements of $G$ change phase of $\hat{\Phi}({\bf k})$ by
$\pi$, and, consequently, the symmetry group of superconducting state is
expressed in the form $G(G')\times R$, where $G'$ is an invariant
subgroup of $G$ [$^{75}$]. From this follows that the modulus $|\hat
\Phi({\bf k})|$ still has the full crystal symmetry and the
superconducting state is nonmagnetic. In both cases the order
parameters are {\it one-component} and their space variations are
described by one complex amplitude $\Psi({\bf r})$. For many purposes
it is sufficient to write the free energy density up to the forth
order in powers of $\Psi({\bf r})$. As a result one obtains the usual
Ginzburg-Landau functional:
\begin{eqnarray}
F & = & \mbox{}-\alpha(T) |\Psi|^2 + \beta |\Psi|^4 + K |D_k \Psi|^2 +
\frac{h^2}{8 \pi} - \frac{{\bf B H}}{4 \pi}\ ,        \label{GL1} \\
& & \alpha(T) = \alpha'\left(1-\frac{T}{T_c}\right),\qquad
D_k=\partial_k-i\,\frac{2e}{\hbar c} A_k\ ,\qquad k=x,y,z\ ,
\nonumber
\end{eqnarray}
where $\alpha$, $\beta$, $K$ are positive phenomenological constants.
Magnetic induction $\bf B$ is a spatial average of the
microscopic magnetic field: $\bf B = \langle h(r)\rangle$ ($\bf h={\rm
rot}A$). Note that for uniaxial (biaxial) crystals, the coefficient $K$
should be replaced by a uniaxial (biaxial) second rank tensor.

     Thus the magnetic properties of one-component unconventional
superconductors and conventional superconductors are similar in the
framework of GL theory. For this reason we will be interested only in
multicomponent unconventional superconductors which correspond to the
multidimensional irreducible representations of $G$.

     In our review we will consider magnetic properties of
unconventional superconductors mostly on the example of
two-dimensional irreducible
representations $E_1$ of group $D_{6h}$ with basis functions
$\hat\Phi_x({\bf k})=k_xk_z$, $\hat\Phi_y({\bf k})=k_yk_z$ in the case
of even order parameter $E_{1g}$, or $\hat\Phi_x({\bf k})=\hat z k_x$,
$\hat\Phi_y({\bf k})=\hat z k_y$ for odd representation $E_{1u}$.
Vector $\eta_i$ in (3) has two components $\eta_x$ and $\eta_y$, and
the GL functional can be written as follows:
\begin{eqnarray}
F & = & \mbox{}-\alpha(T) (\bg{\eta}^*\!\!\cdot\!\bg{\eta}) +\beta_1
(\bg{\eta}^{*}\!\!\cdot\!\bg{\eta})^{2} + \beta_2 |\bg{\eta}\! \cdot
\!\bg{\eta}|^2 + K_1 D^*_i\eta^*_j D^{_{}}_i\eta^{_{}}_j +
K_2 D^*_i\eta^*_i D^{_{}}_j\eta^{_{}}_j +                  \nonumber\\
 & & \mbox{}+K_3 D^*_i\eta ^*_j D^{_{}}_j\eta^{_{}}_i +
K_4 D^*_z\eta^*_i D^{_{}}_z\eta^{_{}}_i
+ \frac{h^2}{8 \pi} - \frac{{\bf B H}}{4 \pi}\ , \qquad i,j=x,y\ ,
\label{GL2}   \\
 & & \hat{\Delta}({\bf k},{\bf r}) = \eta_x({\bf r})\hat{\Phi}_x
({\bf k}) + \eta_y({\bf r})\hat{\Phi}_y ({\bf k})\ .
\nonumber
\end{eqnarray}
Our choice is explained by two reasons.

\noindent
(i) The symmetry group $D_{6h}$ of functional (\ref{GL2}) is
the same as of the normal state of
the heavy fermion superconductor UPt$_3$. It is the $E_1$
irreducible representation that is involved in many models of
interpretation of the complicated phase diagram of UPt$_3$
[$^{24,28,45,74}$].

\noindent
(ii) The 2D model (\ref{GL2}) is, in a sense, the simplest example of
multicomponent order parameter which possesses the nontrivial
properties of the mixed state. Many properties of the GL functional
(\ref{GL2}) may be straightforwardly transferred on another more
complicated models of superconducting pairing in UPt$_3$
[$^{19,53,83}$].

     The symmetry of the multicomponent order parameter is determined by
the coefficients $\eta_i$ in expansion (3). Therefore, in order to
find the residual symmetry of the superconducting state, the
corresponding GL functional should be minimized. The complete list of
the possible residual symmetries for the unconventional
superconductivity in heavy fermions is presented in [$^{22,65}$]. The
residual symmetries of the uniform superconducting state in $E_1$
model will be discussed in Sec.\ 1.3.

\subsection{Magnetic Properties of Usual Superconductors}

     One of fundamental properties of superconductors is the expulsion
of magnetic field from a sample (the Meisner effect). Depending on the
material, the Meisner effect can be either complete, as in type-I
superconductors, or partial, as in type-II superconductors. In the latter case
magnetic field is dispersed through the material in the form of
flux lines or vortices, each of them containing a unit of magnetic
flux $\phi_0= h c/2|e|=2.07\cdot 10^{-7}$ G$\cdot$cm$^2$.
Note that heavy
fermion superconductors demonstrate properties of the type-II
superconductors.

     Having in mind the discussion of the mixed state in
unconventional superconductors, we survey the main magnetic properties
of usual type-II superconductors with $s$-pairing, which can be
understood on the basis of the GL functional (\ref{GL1}). Varying
energy (\ref{GL1}) with respect to $\delta\Psi^*$ and $\delta\bf A$,
one gets the GL equations:
\begin{equation}
\mbox{} - \alpha\Psi + 2\beta |\Psi|^2\Psi - K D_iD_i\Psi = 0\ ,
\label{GLeq1}
\end{equation}
$$
(\bg{\nabla}\times(\bg{\nabla}\times {\bf A})) =
\frac{4\pi}{c}\,{\bf j}\ , \qquad
{\bf j}=\frac{2e}{i\hbar}\,K\,(\Psi^*\bg{\nabla}\Psi -
\Psi\bg{\nabla}\Psi^*) -
\frac{8e^2K}{\hbar^2c}\,|\Psi|^2{\bf A}\ .
$$
According to the classical results of Abrikosov [$^{1}$], magnetic
properties of type-II superconductors are quite different in
different parts of the $H$--$T$ phase diagram.

     For magnetic fields smaller than the {\it lower critical field}
$H_{c1}$, circular surface currents screen the external field and
magnetic flux does not penetrate inside the sample. The amplitude of
superconducting order parameter is fixed: $|\Psi_0|^2=\alpha/2\beta$,
whereas the phase of $\Psi_0$ is arbitrary. The gain of the energy in
superconducting state defines the {\it thermodynamical critical field} $H_c$:
$F_n-F_s=H_c^2/8\pi$.

     For $H>H_{c1}$ the flux begins to penetrate inside the material in
the form of vortices. A single vortex line has an axially symmetric
structure:
\begin{equation}
\Psi({\bf r})=\Psi(r)\exp(-i\varphi)\ ,\qquad {\bf h}({\bf r})=
h(r)\,\hat{\bf z}\ , \label{oneqv}
\end{equation}
($r$ and $\varphi$ are the polar coordinates) and carries one quanta
of magnetic flux [$^{16,59}$]. The amplitude $\Psi(r)$ is suppressed
to zero at the vortex center. It restores to the equilibrium value
$\Psi_0$ at a radius of the vortex core, which is of the order of the
coherence length
\begin{equation}
\xi = \left(\frac{K}{\alpha}\right)^{1/2}.  \label{xi}
\end{equation}
The magnetic field is maximal at the vortex center and approximately
equal to $2H_{c1}$. The supercurrents (which are beyond the vortex
core) screen the magnetic field and $h(r)$ tends to zero outside
the screening region which is of the order of the London penetration
depth
\begin{equation}
\lambda = \left(\frac{\hbar^2c^2\beta}
{16\pi e^2\alpha K}\right)^{1/2}.
\end{equation}
The ratio of the penetration depth to the coherence length defines a
dimensionless temperature independent (near $T_c$) GL parameter $\kappa$, whose
value determines the type of superconductor. For type-II
superconductors $\kappa > 1/\sqrt{2}$. It is the screening supercurrents
$j_s= c\,\phi_0/(8 \pi^2\lambda^2 r)$, $\xi\ll r\ll\lambda$ which bring for
large $\kappa$ the main logarithmic contribution to the vortex energy
per unit length or ``flux line tension:''
\begin{equation}
\varepsilon_L = \frac{\phi_0^2}{(4\pi \lambda)^2} \left(\ln
\frac{\lambda}{\xi}+ \epsilon \right),          \label{tension}
\end{equation}
where the small numerical constant $\epsilon \approx 0.08$
is due to the vortex core energy.
The lower critical field is defined by the expression:
\begin{equation}
H_{c1} = \frac{4\pi}{\phi_0 }\cdot\varepsilon_L =
\frac{\phi_0}{4\pi \lambda^2} \left(\ln\frac{\lambda}{\xi}+ \epsilon
\right).           \label{Hc1}
\end{equation}
When the intervortex distance $r_L$ is larger than $\lambda$,
it is sufficient to take into account interaction only between
nearest neighbors through pair potential $U= \phi_0/4
\pi\!\cdot\!h(r_L)$. In the intermediate region
$\xi\ll r_L \le \lambda$ ($H>H_c$) vortex lines form a dense lattice. Single
vortex interacts with all neighbors in the region of the area
$\lambda^2$.

     When increasing field reduces the distance between vortices to the
order of $\xi$, the superconductivity is destroyed.
To calculate this the {\it upper critical field} $H_{c2}$, one should
consider the GL equation (\ref{GLeq1}) neglecting the diamagnetic
field generated by superconducting currents and the nonlinear term,
which are small near $H_{c2}$. Thus we derive the {\it linearized GL
equation}:
\begin{equation}
\alpha\Psi = -K\left(\bg{\nabla} - i\,\frac{2e}{\hbar c}{\bf A}
\right)^{2}\Psi\ ,             \label{Shred}
\end{equation}
which formally coincides with the Schr\"odinger equation for an
electron in a magnetic field. This equation is known to have a discrete
set of Landau levels. Using the analogy with the Landau level
quantization, we obtain a set of eigen (critical) fields:
\begin{equation}
H_{n}=\frac{\hbar c\alpha}{2|e| K(2n+1)} =
\frac{\phi_0}{2\pi\xi^2}\cdot\frac{1}{2n+1}\ .
\end{equation}
The upper critical field corresponds to the highest $H_{n}$ and,
therefore, to the lowest Landau level with $n=0$:
\begin{equation}
H_{c2}= \frac{\phi_0}{2\pi \xi^2}\ .
\end{equation}
Vortices form a regular lattice of equilateral triangles at the entire
field region between $H_{c1}$ and $H_{c2}$.

\subsection{New Features of Multi-Component Superconductors}

     As we have seen in previous Subsection, the energy functional
(\ref{GL1}) leads to the particular magnetic properties.
These properties will be the same (at least near $T_c$) for all
one-component superconductors, since they are also described by
functional (\ref{GL1}) or its anisotropic modifications. Therefore we are
interested mainly in superconductors described by a multicomponent
order parameter.

     The dimensionality of the order parameter is that of the
irreducible representation to which it belongs. Additional degrees of
freedom always correspond to the states of the Cooper pairs degenerated
with respect to the direction of the internal angular momentum, it
does not matter whether of spin or orbital origin. For example, in the
case of the 2D order parameter, the degeneracy exists between the
states with ``up'' or ``down'' direction of the angular momentum $i
(\bg{\eta}^*\!\times\!\bg{\eta})$.

     In the case of GL energy (\ref{GL2}), the equilibrium order
parameter satisfies the following nonlinear equations:
\begin{equation}
\alpha\eta_i-2\beta_1\eta_i^{_{}}(\eta_j^*\eta_j^{_{}}) - 2\beta_2 \eta_i^*
(\eta_j^{_{}}\eta_j^{_{}}) + K_1 D_j D_j\eta_i + K_2 D_i D_j\eta_j +
K_3 D_j D_i\eta_j + K_4 D_z D_z\eta_i\,=\,0,            \label{GLeq2}
\end{equation}
$$
j_i=-\frac{4e}{\hbar}\,{\rm Im}
\left[K_1\eta_j^*D_i^{_{}}\eta_j^{_{}} + K_2 \eta_i^*D_j^{_{}}\eta_j^{_{}} +
K_3 \eta_j^*D_j^{_{}}\eta_i^{_{}} + K_4 \delta_{iz}\eta_j^*D_z^{_{}}
\eta_j^{_{}}\right].
$$

     New properties in comparison with ordinary superconductors
appear already without magnetic field. Depending on the sign of
$\beta_2$, the resulting superconducting phases are:
\begin{eqnarray}
\hat{\Delta}({\bf k}) = \left(\frac{\alpha}{4\beta_1}\right)^{1/2}
 (\hat{\bf\Phi}_1({\bf k})+i\hat{\bf\Phi}_2 ({\bf k}))\,,\, &
(|\hat{\bf\Phi}_1| = |\hat{\bf\Phi}_2|,\,\hat{\bf\Phi}_1\perp
\hat{\bf\Phi}_2)\,,\  &\beta_2>0\,,     \label{compl} \\
\hat{\Delta}({\bf k}) = \left(\frac{\alpha}{2\beta_{12}}
\right)^{1/2}\hat{\bf\Phi}({\bf k})\,,\qquad\qquad\quad &
(\hat{\bf\Phi}=\hat{\bf\Phi}^*)\,,\qquad\qquad\quad\ \ &
\beta_2<0\,.              \label{real}
\end{eqnarray}
To be consistent, the energy should be bounded from below for each phase
(\ref{compl}) and (\ref{real}) what leads to the positive definiteness of
the forth order terms:
\begin{equation}
\beta_1>0\ ,\qquad \beta_1+\beta_2=\beta_{12}>0\ .
\end{equation}

     For $\beta_2>0$ the ground state of the GL functional (\ref{GL2})
has two-fold degeneracy with respect to the appearance of either
$\hat{\bf\Phi}_+ = \hat{\bf\Phi}_1 + i\hat{\bf\Phi}_2$ or
$\hat{\bf\Phi}_-=\hat{\bf\Phi}_1 - i\hat{\bf\Phi}_2$ phases. Both
phases are conjugated with respect to the time inversion. Physically,
these phases correspond to the superconducting states with definite
values of the projection of internal angular momentum of Cooper pairs like in
$^3$He-A.
Symmetry group of
$\hat{\bf\Phi}_\pm({\bf k})$ phase includes arbitrary rotations about
$\hat{\bf z}$ combined with multiplication by the phase factor
$\exp(\mp i\varphi)\hat{L}_{\varphi}$, and rotations by an angle of
$\pi$ about $\hat{\bf x}$ combined with the time reversal $RU_{2x}$.
Discrete degeneracy of the energy minimum leads to the two types of
superconducting domains.

     For $\beta_2<0$ the order parameter is real up to the complex
factor. The symmetry group of the superconducting state in this case
is $D_2(C_2)\times R$. The direction of the vector $\hat{\bf\Phi}({\bf
k})$ is fixed by the sixth order term $\eta_+^{*3}\eta_-^3 +
\eta_-^{*3} \eta_+^3$ in the GL functional ($\eta_\pm = \eta_x \pm
i \eta_y$). We have not included this term in (\ref{GL2}) since it is
small near $T_c$ in comparison with the other terms. In this case the
ground state possesses continuous degeneracy with respect to the
direction of $\hat{\bf\Phi}$. Therefore smooth texture of the order
parameter $\hat{\Delta}({\bf k},{\bf r}) = {\bf n}({\bf r}) \cdot
\hat{\bf\Phi}({\bf k})$ should be expected in real samples rather
than domains.

     Derivation of the free energy functional (\ref{GL2}) in the
framework of the Landau theory requires as a necessary condition
positiveness of the gradient terms without any magnetic field (i.e.,
if ${\bf A(r)}\equiv 0$), which leads to:
\begin{equation}
K_1>0\ ,\qquad K_4>0\ ,\qquad K_{123}=K_1+K_2+K_3>0\ .    \label{cond}
\end{equation}
However within this range of parameters it is possible the magnetic
instability of the uniform superconducting state [$^{54}$], which
results in a spatially varying order parameter in zero applied field
[$^{81}$]. In fact exact restrictions connected to the existence of
the lower energy boundary for such laminar phases do not differ
strongly from the above necessary conditions [$^{81}$]. Therefore in
the following consideration we impose only conditions (\ref{cond}). It
will be useful to introduce dimensionless parameters
$C=(K_2+K_3)/2K_1$, $D=(K_2-K_3)/2K_1$. Then, according to
(\ref{cond}), we have $1+2 C>0$.

     Neglecting by influence of the order parameter $\hat{\Delta}$ on
the pairing potential, one can obtain in the weak-coupling approximation
the following relations between phenomenological parameters for
$E_{1g}$ representation [$^{22}$]: $\beta_2=0.5\beta_1$,
$K_1=K_2=K_3$. Note that the approximate particle-hole symmetry of
Bogoljubov equations near the Fermi surface results in a relatively
small value of $D \sim (T_c/E_F)^2$ [$^{22,14,55}$].

     Returning to the magnetic properties of unconventional
superconductors, we could expect the same qualitative features in
their behavior, namely, penetration of separate vortices inside the
material at fields higher than $H_{c1}$, formation of the vortex
lattice, which becomes denser with increasing external field, and
destruction of superconductivity at $H=H_{c2}$ when the distance
between vortices becomes of the order of $\xi$.

     However, the GL equations (\ref{GLeq2}) for the 2D order
parameter have a more complicated structure than those for the
$s$-wave superconductor (\ref{GLeq1}). This is why new properties of
the mixed state of unconventional superconductor are expected, and the
well established results for usual superconductors should be
reexamined in the multicomponent case.

     The structure of the mixed state of a multicomponent
superconductor is considered in the following Sections of this
review. Here we give a short outline of the problems to be discussed.
\smallskip

\noindent 1. The multicomponent structure of the order parameter and
two different types of ground state lead to the possibility of the
existence of different types of vortices carrying an integer flux,
which are discussed in Sec.\ 2. Unlike the case of usual
superconductors, the vortices can have a nonsingular structure of the
core and be nonaxisymmetric. Since in reality only the most
energetically favorable vortices exist, one should calculate the
energy for each type of vortices and choose the vortex with the lowest
energy. Note that the favorability of single-quantum vortices over
multiple-quanta vortices is also questionable.
\smallskip

\noindent 2. In order to calculate the upper critical field
$H_{c2}(T)$ for unconventional superconductors, the multicomponent
linearized GL equations have to be solved. The puzzling feature of
solutions of such equations described in Sec.\ 3 is that all
components of the order parameter participate in the formation of the
Abrikosov vortex lattice, and, hence, the internal structure of the
mixed state near $H_{c2}$ is rather complicated. The main question to
be considered in this context in Sec.\ 4 is: whether the lattice has
the regular triangular form as for $s$-wave superconductors or it is
distorted. Another surprising fact is the existence of sevral
superconducting phases from different Landau levels with close values
of critical fields. It
will be shown in Sec.\ 5 that the admixture of such phases
inevitably leads to structural phase transitions in the Abrikosov
lattice below $H_{c2}$.

In addition to the review of already published works we have included
new original results in Secs.\ 3.5 and 4.3.
\smallskip

       In fact, depending on the parameters in multicomponent GL
equations, both nontrivial vortices near $H_{c1}$ and structural phase
transitions in the Abrikosov lattice near $H_{c2}$ may occur. The
crucial open question of the mixed state theory of the
multicomponent superconductors is relation of these two limiting
cases:

\noindent --- How are nonaxisymmetric vortices
placed in the lattice near $H_{c1}$?

\noindent --- What is the behavior of such vortex lattice with the
increase of magnetic field?

\noindent --- What kinds of phase transitions
can occur in the intermediate region
between $H_{c1}$ and $H_{c2}$ in unconventional superconductors?

 \section{Vortices in Unconventional Superconductors}

     The vortices in a superfluid liquid with the multicomponent order
parameter were intensively studied during the last two decades in
connection with rotating superfluid $^3$He (for review see
[$^{60}$]).
It was shown that the response of such a liquid to rotation is
formation of topologically stable vortices of the order parameter. At
large distances from the vortex core the slow space variation of
the order parameter $\hat{\Delta}({\bf k},{\bf r})$ can be
described as a long-range texture of the uniform order parameter
$\hat{\Delta}({\bf k})$ which is fixed by the forth order terms in
the GL functional. (For $^3$He it is the space variation of A or B
phase.) However,
in the vicinity of the vortex core the bulk phase is destroyed and
all components of
the order parameter have the same order of magnitude.
We define the characteristic size
of this region as $\tilde \xi$.

     The analog of the superfluid liquid rotation is the response of
the superconductor to the external magnetic field. Before translating
the results of superfluid $^3$He theory on the case of the multicomponent
superconductors,
we should stress that there is a feature which makes the difference  between
superconducting and superfluid vortices: the first ones have the
finite size $\lambda$ due to the screening effect of the currents,
whereas the size of superfluid vortices is infinite.

     Therefore, three length scales enter the problem of vortices in
multicomponent superconductors: the vortex size (the London
penetration depth) $\lambda$, the radius of the vortex ``hard'' core
(the coherence length) $\xi$, and the radius of the region around the
vortex core $\tilde \xi$ where all ${\bf k}$-components of the order
parameter are mixed. Note, however, that such a stratification of the
vortex on the characteristic scales has physical sense only under
the condition that $\lambda \gg \tilde \xi \gg \xi$ where the values
$\lambda$, $\tilde \xi$, $\xi$ are defined by the parameters of multicomponent
GL
functional. Having in mind this case, we will discuss magnetic
vortices in the two-component model (\ref{GL2}), which have been
studied in most details. We will assume also that the magnetic field
is parallel to the principal crystal axis.

     The properties of vortices and the scales of $\lambda$, $\tilde
\xi$, $\xi$ strongly depend on the bulk superconducting state.
As we have already mentioned in Sec.\ 1.3 there
are two stable minimums of the GL functional (\ref{GL2}) describing
homogeneous state for different signs of $\beta_2$. There is either a
complex order parameter (broken time-reversal phase)
$\hat{\Delta}({\bf k}) \sim \hat{\bf\Phi}_1 + i\hat{\bf\Phi}_2$,
($\beta_2>0$) or a real one (vector phase) $\hat{\Delta}({\bf
k})\sim \hat{\bf\Phi}$, ($\beta_2<0$). We will consider both cases
separately in Secs.\ 2.1 and 2.2.

\subsection{Vortices in the Broken Time-Reversal Phase}

     To begin with the discussion of vortices in the broken
time-reversal phase, note that the ground state is doubly degenerate
with respect to the direction of the angular momentum of Cooper pairs. Due
to this discrete degeneracy, the domain structure (with phases
$\hat{\bf\Phi}_+({\bf k})$, $\hat{\bf\Phi}_-({\bf k})$) in the ground
state of the two-component superconductor is expected [$^{75}$]. Thus,
the vortices originating from both phases can exist inside the
superconductor. Note also that the domain structure itself contributes
to the magnetic properties of the unconventional superconductors
because of two important effects: (i) the internal (``up'' and
``down'') magnetization of Cooper pairs owning to their orbital
momentum (this bulk magnetization is assumed to be small due the
microscopic electron-hole symmetry [$^{76}$]; (ii) the magnetization
owning to persistent surface currents which circulate around domains
and originate due to the inhomogeneity of the order parameter (on the
scale $\tilde\xi$) near the domain boundary (these currents are due to
boundary effects of the bulk magnetization [$^{76}$]). The low field
magnetic response of unconventional superconductors due to the domain
structure was considered in detail in [$^{63,64,65}$].

     Note that an external magnetic field lifts the
$\hat{\bf\Phi}_\pm({\bf k})$ degeneracy and makes one sort of domains
more favorable. Therefore, the ground (vortex) state of unconventional
superconductor in an external field is rather complicated and history
(field or zero-field cooling experiments) dependent. In this review we
will neglect for simplicity the effects of domain- and
internal-magnetization and will assume that the vortices exist in the
bulk of domains (the latter means that the domain sizes are larger
than the penetration length). We will discuss the bulk vortices in
both phases $\hat{\bf\Phi}_+({\bf k})$ and $\hat{\bf\Phi}_-({\bf k})$
simultaneously using ``$\pm$'' subscript.

\subsubsection{Global vortex structure}

     The long range structure of $n$-quantum
vortex is given by:
\begin{equation}
\hat{\Delta}_\pm({\bf k},{\bf r}) = \Psi_\pm({\bf r})
(\hat{\bf\Phi}_1({\bf k})\pm i\hat{\bf\Phi}_2({\bf k}))\ ,
                                                        \label{vort}
\end{equation}
where phase of $\Psi_\pm({\bf r})$ varies on $2\pi n$ about the
vortex center. Substituting (\ref{vort}) into (\ref{GLeq2}), we obtain
the effective GL equation for the space variation of $\Psi_\pm({\bf
r})$:
\begin{equation}
\left(\mbox{}-\alpha \mp \frac{2e}{\hbar c}\,h({\bf r}) DK_1\right)
\Psi_\pm + 4\beta_1|\Psi_\pm|^2\Psi_\pm - K_1(1+C)D_i D_i\Psi_\pm=0\ .
\label{GLeq1'}
\end{equation}
The term proportional to $D$ is responsible for the above mentioned
interaction of the angular momentum of Cooper pairs with local
magnetic field. If we neglect by this term (see Sec.\ 1.3), Eq.\
(\ref{GLeq1'}) becomes the same as the usual GL equation (\ref{GLeq1}) for
$s$-wave superconductor. Then by analogy we find the size of the
vortex (penetration depth) in the time-reversal breaking phase:
\begin{equation}
\lambda_t = \left(\frac{\hbar^2c^2\beta_1}{16 \pi e^2 \alpha
K_1(1+C)}\right)^{1/2}.
\end{equation}
We can also define the coherence length which corresponds to the size
of the vortex core:
\begin{equation}
\xi_t = \left(\frac{K_1(1+C)}{\alpha}\right)^{1/2}.
\end{equation}
However, this length has a limited physical sense for small $\beta_2$
because in the region $r \le \tilde\xi_t$ other components of the order
parameter are admixed, and Eq.\ (\ref{GLeq1'}) does not hold. In this
region the gradient energy due to the texture of $\hat{\bf\Phi}_\pm$
\begin{equation}
F_{grad}\sim \frac{K_1(1+C)}{\tilde\xi_t^2}\,F_{bulk}
\end{equation}
becomes comparable to the contribution to the uniform part of the GL
energy due to other admixed components of the order parameter:
\begin{equation}
\delta F_{bulk} \sim \frac{\beta_2}{\beta_1}\,F_{bulk}\ .
\end{equation}
Equating $\delta F_{bulk}$ to $F_{grad}$, we obtain the characteristic
scale $\tilde\xi_t$:
\begin{equation}
\tilde\xi_t =\xi_t \left( \frac{\beta_1}{\beta_2}\right)^{1/2}.
\label {txi}
\end{equation}
The same estimation is valid also for the width of the domain wall between
$\hat{\bf\Phi}_+$ and $\hat{\bf\Phi}_-$ phases [$^{25,64}$].

By the analogy with usual $s$-wave superconductor, we can conclude
that the one-quantum vortex has the smallest energy and $H_{c1}$ is
defined by formulas (\ref{oneqv}) and (\ref{Hc1}) with a formal
substitution $\xi \rightarrow \tilde\xi_t$,  $\epsilon \rightarrow
\tilde \epsilon_t$, where $\tilde \epsilon_t$ is the energy of the
vortex core in the region $r < \tilde\xi_t$.
In the region $r \gg \tilde\xi_t$ the vortex (\ref{vort}) is
described by the same function (\ref{oneqv}): $\Psi_{\pm}({\bf
r})=\Psi(r)e^{-i\varphi}$ (with substitution $\xi_t$, $\lambda_t$)
like a usual Abrikosov vortex.

     These statements are valid provided $\lambda_t\gg \tilde\xi_t$,
which is the case for not very small positive $\beta_2$. In
particular, we have assumed that the main vortex energy is
concentrated in the screening currents region $\lambda_t > r >
\tilde\xi_t$, what means that $\ln(\lambda_t /\tilde\xi_t) \gg
\tilde\epsilon_t$.

\subsubsection{Vortex core structure}

     The investigation of the vortex core structure at $\beta_2>0$ was
done in [$^{70,71,6}$]. Our description of the vortex structure
is based on the results of these papers. We will use the analytical
arguments of [$^{6}$] to illustrate the appearance of vortices
with nontrivial core structure and then present the results of the
numerical calculations of [$^{70}$] which were done in a broad range
of parameters.

     In the region $r\le\tilde\xi_t$ all components of the order
parameter are mixed:
\begin{equation}
\hat{\Delta}_{\pm}({\bf k},{\bf r}) =
(\hat{\bf\Phi}_1({\bf k}) \pm i\hat{\bf\Phi}_2({\bf k}))
\Psi_{\pm}({\bf r}) +
(\hat{\bf\Phi}_1({\bf k})\mp i\hat{\bf\Phi}_2
({\bf k})) X_{\mp}({\bf r})\ .                      \label{axvort}
\end{equation}
Functions $\Psi_\pm({\bf r})$, $X_\mp({\bf r})$ can be determined from
solution of the GL equations (\ref{GLeq2}). Since at $r\gg\xi_t$ vortex
(\ref{axvort}) should gradually turn into (\ref{vort}), we have at the
large distances $\Psi_\pm({\bf r}) \rightarrow \Psi(r)e^{-i\varphi}$
and $X_\mp({\bf r})\rightarrow 0$.
\medskip

\noindent {\bf Axisymmetric vortices}
\medskip

     In general, there exist different solutions of Eqs.\
(\ref{GLeq2}) obeying the boundary conditions
$\Psi_\pm|_{r\rightarrow\infty}=\sqrt{\alpha/4\beta_1}\,e^{-i\varphi}$,
$X_\mp|_{r\rightarrow\infty} = 0$. One
of the extremums of GL functional always corresponds to the {\it ``most
symmetric vortex,''\/} which possesses all symmetries possible
for line defects in condensed matter [$^{60}$]. According to
[$^{60}$] the maximal symmetry group of a line defect includes
continuous rotations about its axis with multiplications by a phase
factor, the reflection $\hat{\sigma}_h$ in the perpendicular plane,
and the combined symmetry $RU_2$. The last two discrete symmetries
have no effect on vortex classification in the model of 2D
superconducting order parameter.

     The generator of continuous subgroup and corresponding condition
of invariance have form:
\begin{equation}
\hat{Q}\,\hat{\Delta}^a_\pm({\bf k},{\bf r}) = 0\ ,
\qquad \hat{Q} = \hat{L}_z - q \hat{I}\ .            \label{cndinv}
\end{equation}
Here $\hat I$ is the generator of $U(1)$ group, and $q$ is an integer
which has the meaning of the total angular momentum eigenvalue of the
vortex. The symmetry group of the axisymmetric vortex is
$\{e^{-iq\varphi} \hat{L}_{\varphi}, RU_2, \hat{\sigma}_h \}$. The
solutions of Eq.\ (\ref{cndinv}) for the case of one-quantum vortices
in the phases $\hat{\bf\Phi}_\pm$ are
\begin{equation}
\begin{array}{lll}
\Psi_+^a({\bf r}) = \Psi_+(r)\,e^{-i\varphi}\ , \quad
 & X_-^a({\bf r}) = X_-(r)\,e^{i\varphi}\ , \qquad\quad & q=0 \ ,  \\
\Psi_-^a({\bf r}) = \Psi_-(r)\,e^{-i\varphi}\ ,
 & X_+^a({\bf r}) = X_+(r)\,e^{-3i\varphi}\ ,   & q=-2 \ .
\end{array}
\label{axcomp}
\end{equation}
Functions $\Psi_\pm(r)$ and $X_\mp(r)$ vanish at $r = 0$ making a
``hard'' core of the axisymmetric one-quantum vortex. At large
distances $|\Psi_\pm|^2\approx \alpha/4\beta_1$, and amplitude $X_\mp$
is small. Linearizing Eq.\ (\ref{GLeq2}) in $X_\mp$ in this region,
one comes to the following equation:
\begin{equation}
\left( \mbox{} - \alpha + 4(\beta_1+2\beta_2)|\Psi_\pm|^2 -
K_1(1+C)(D_x^2+D_y^2)\right)X_\mp = K_1C(D_x \pm i D_y)^2 \Psi_\pm \ .
                         \label{well}
\end{equation}
Substituting $\Psi_\pm = \sqrt{\alpha/4\beta_1}\,e^{-i\varphi}$ in
(\ref{well}), one obtains:
\begin{eqnarray}
X_-^a({\bf r}) & \approx &  \mbox{}-\frac{C}{2(1+C)}
\left(\frac{\tilde\xi_t}{r}\right)^2
\left(\frac{\alpha}{4\beta_1}\right)^{1/2} e^{i\varphi}\ , \nonumber \\
X_+^a({\bf r}) & \approx &  \quad\ \frac{3C}{2(1+C)}
\left(\frac{\tilde\xi_t}{r}\right)^2
\left(\frac{\alpha}{4\beta_1}\right)^{1/2} e^{-3i\varphi}\ .
\label{estim}
\end{eqnarray}
 From (\ref{estim}) one can easily estimate the region where
$X_\mp^a$ becomes comparable to $\Psi_\pm^a$. For $C\ge 1$ its sizes
coincide with $\tilde\xi_t$. In order to find behavior of
$\Psi_\pm(r)$ and $X_\mp(r)$ at $r \rightarrow 0$, one should solve
both GL equations (\ref{GLeq2}) simultaneously. The admixture of the other
components changes the usual assimptotic $\Psi_\pm(r)\sim r$ of
superconducting
amplitude at the vortex core.
\medskip

\noindent {\bf Nonaxisymmetric vortices}
\medskip

     Although the axially symmetric vortex solution is always extremum, it is
not necessarily the absolute minimum of the GL functional. Obviously, it is
the minimum as $\tilde\xi_t \rightarrow 0$ ($\beta_2\gg \beta_1$),
when $\Psi_\pm^a({\bf r})$ is defined by Eq.\ (\ref{well}) almost
throughout the entire space. Functions $\Psi^a_\pm({\bf r})$ are
solutions of (\ref{well}) in all space also for $C=0$ when the
additional amplitude in axisymmetric vortex $X_\mp^a({\bf r}) \equiv
0$. Due to this feature the case of $C=0$ is a convinient starting point
for consideration of nonaxisymmetric vortices.

In this case Eq.\ (\ref{well}) becomes uniform and takes form
of the eigenvalue problem for the ordinary Schr\"odinger equation on
$X_\mp$ in a potential well with $U(r) = - \alpha + 4 (\beta_1 +
2\beta_2) \Psi^2(r)$. A two-dimensional well always possesses a bound
state, which is placed on some finite energy below the value of potential at
infinity $U(r)|_{r\rightarrow\infty} = \alpha\beta_2 / 2\beta_1 > 0$.
Therefore at
$C=0$ the lowest energy level of (\ref{well}) should pass through 0 as
$\beta_2$ decreases. This means that the vortex $\hat{\Delta}
^a_\pm ({\bf k},{\bf r}) = \Psi^a_\pm({\bf r})\hat{\bf\Phi}_\pm({\bf
k})$ becomes unstable towards perturbation of the form $X^0(r)
\hat{\bf\Phi}_\mp({\bf k})$. Here $X^0(r)$ is a real axisymmetric
function exponentially decreasing at $r\gg\tilde\xi_t$ and with maximum at
$r=0$,
which corresponds to the lowest level in the well.

     The critical value of $\beta = \beta_2/\beta_1$ corresponding to
such an instability in the vortex core is $\beta_c = 0.24$ according
to [$^{70}$]. Using trial functions Barash and Mel'nikov [$^{6}$]
have found a slightly different value: $0.37$. However, subsequent
numerical investigations of one-component
odd-parity model [$^{46}$], which reduces in a magnetic field to model
(\ref{GL2}) with $C=0$, have suggested the same value for $\beta_c$ as in
[$^{70}$]. The order parameter amplitude at the center of such {\it
nonsingular} vortex grows as $(\beta_c - \beta)^{1/2}$.

     When $C$ is exactly equal to zero, sum $|\Psi_\pm({\bf r})|^2 +
|X^0(r)|^2$ depends only on $r$. This leads to axial symmetry of
the whole vortex. Such a symmetry of the nonsingular vortex, however, is
``accidental'' and disappears for any small value $C$.

     For small $C$ solution of Eq.\ (\ref{well}) is the
superposition $X_\mp({\bf r}) = X_\mp^a({\bf r}) + X^0(r)$.
Consequently, $X_\mp({\bf r})$ is not already invariant under
arbitrary rotations and axisymmetry is spontaneously broken. Moreover,
using (\ref{axcomp}) we come to the following conclusions about
symmetry of the nonsingular vortices:
\smallskip

\noindent
(i) Vortices originating in $\hat{\bf\Phi}_-$ state are
one-quantum, axisymmetric and singular for $\beta_2/\beta_1 >
\beta_c^-(C)$. For $\beta_2/\beta_1<\beta_c^-(C)$ they have {\it
nonaxisymmetric triangular} structure (Fig.\ 2a). The residual
symmetry group is: $\{e^{4\pi i/3}\hat{L}_{2\pi /3}, \hat\sigma_h, R
\hat\sigma_v\}$, where $\hat\sigma_v$ means the reflections with
respect to the triangular medians.

\noindent
(ii) One-quantum vortices in $\hat{\bf\Phi}_+$ state have axisymmetric
singular structure for $\beta_2/\beta_1 > \beta_c^+(C)$. For $\beta_2/
\beta_1 < \beta_c^+(C)$ vortex cores are of {\it nonaxisymmetric
``crescent''\/} shape (see Fig.\ 2b). The residual vortex symmetry is of the
vector type
and contains only two elements: $\{ \hat\sigma_h, R\hat\sigma_v \}$,
where $\hat\sigma_v$ is the reflection in the symmetry plane of ``crescent.''
\smallskip

     The boundary of the axisymmetric vortex instability
$\beta_c^\pm(C)$ was found by Tokuyasu, Hess and Sauls [$^{70}$]
numerically (see Fig.\ 3). The results of [$^{6}$] at small $C$
qualitatively coincide with [$^{70}$].

     For topological reasons, only the vortices with an integer number
of flux quanta can exist in the bulk of time-reversal breaking phase.
However, vortices carrying fractional number of magnetic flux quanta are
known on a domain boundary between phases
$\hat{\bf\Phi}_+$ and $\hat{\bf{\Phi}}_-$ [$^{65}$]. Moreover, two
fractional vortices could have a lower energy than a single-quantum vortex.
Starting on this picture, Izumov and Laptev
[$^{25}$] suggested a scenario in which a bulk vortex decays
forming a small domain of the time-conjugated phase and two
fractional vortices which exist on the closed boundary of the new
domain. The loss in the gradient energy due to the domain boundary is
compensated by the gain of energy due to the vortex decay. Note that the
vortex core instability in the phase $\hat{\bf\Phi}_+$ with the
formation of the nonaxisymmetric crescent structure and phase
$\hat{\bf\Phi}_-$ inside the vortex core is an extreme case of the
Izumov and Laptev scenario when the domain size is sufficiently small.

     Developing the analogy with vortices
in $s$-wave superconductor established by Eq.\ (21) we can
estimate vortex core energy $\tilde\epsilon_t$ as
$(\ln \tilde\xi_t /\xi_t + \epsilon)$.
Then for those $\beta_2$, which satisfy condition
$\tilde\xi_t/\xi_t > \lambda_t / \tilde\xi_t$ or
$\beta_2<\beta_1/\sqrt{\kappa}$, the main part of the vortex line
tension is the energy of the vortex core $\tilde\epsilon_t$ rather
than the hydrodynamics part proportional to $\ln
(\lambda_t/\tilde\xi_t)$. In this case the energetic favorability of the
one-quantum vortices in comparison with multi-quantum ones is broken.
For example, the {\it two-quantum axisymmetric vortex} in the phase
$\hat{\bf\Phi}_+$ has according to (\ref{cndinv}) nonvanishing
amplitude $X_-^a({\bf r})$ at the center and therefore smaller core
energy $\tilde\epsilon_t$ than one-quantum vortex. Its existence in the
particular parameter
region was found numerically in [$^{71}$].

     Spontaneous breaking of vortex axisymmetry was known before in
superfluid $^3$He-B. In particular parameter region the core of so
called $v$ vortex is subjected to the {\it ``director type''}
instability [$^{60}$]. As it have been shown above the axial symmetry
breaking in the vortex cores is quite general phenomenon in the case
of the multicomponent order parameter. Ordering of such nonaxisymmetric
vortices in the lattice at $H > H_{c1}$ is still an open question.

\subsection{Vortices in the Vector Phase}

\subsubsection{Global vortex structure}

For $\beta_2<0$, the equilibrium order parameter is $\hat{\Delta}
({\bf k}) = ({\bf n}\cdot\hat{\bf\Phi}({\bf k}))\Psi$, where ${\bf n}$
is a unit vector. Far from the vortex core the long range texture of
the order parameter preserves this form:
\begin{equation}
\hat{\Delta}({\bf k},{\bf r}) = ({\bf n}({\bf r})\cdot
\hat{\bf\Phi}({\bf k}))\Psi({\bf r})\ .                \label{vector}
\end{equation}
Substituting (\ref{vector}) into (\ref{GL2}) and keeping the gradient
terms responsible for Goldstone modes, we have:
\begin{eqnarray}
F & = & \mbox{} - \alpha |\Psi|^2 + \beta_{12}|\Psi|^4 + K_1
D_i^*\Psi^*D_i\Psi + K_{23} (n_iD_i^*\Psi^*)(n_jD_j\Psi) +
                      \nonumber    \\
  &   & \mbox{} + |\Psi|^2 (K_1({\rm rot}\,{\bf n})^2 +
K_{123} ({\rm div}\,{\bf n})^2)\ ,
\end{eqnarray}

     There are two topological reasons for the formation of stable
line defects in the vector state (\ref{vector}): (i) the phase of
$\Psi({\bf r})$ changes on $2\pi n$ in making a complete circuit; (ii)
the vector ${\bf n}({\bf r})$ sweeps the unit circle $m$ times when
moving around the vortex. Therefore, possible defects are
classified by two quantum numbers $(n,m)$, both which are integer or
half-integer (for general aspects of topological classification
see  [$^{49}$]).

     Note, however, that the phase variation about the vortex leads to
supercurrents which screen the magnetic field outside the region of the size
$\lambda$. For this reason the vortex energy has the upper cutoff
parameter $\lambda$ and is therefore finite. The variation of the
vector ${\bf n}({\bf r})$ does not produce screening currents.
Therefore, for $m\neq 0,$ the contribution to the defect energy
$\sim\Delta^2\int (K_1 ({\rm rot}\,{\bf n})^2 + K_{123} ({\rm div}
\,{\bf n})^2)dV$ is logarithmically divergent beyond the core. It
makes the defects with $m\neq 0$ less favorable then the pure phase
vortices with $m=0$ but $n\neq 0$.

     Since we are interested in the vortices with minimal energy, we
will consider only those with uniform distribution of ${\bf n}({\bf
r})$ at large distances. Assuming that ${\bf n}({\bf r})$ is uniform,
we rewrite the effective GL equation for $\Delta ({\bf k}, {\bf r}) =
\Psi({\bf r})({\bf n}\cdot\hat{\bf\Phi}({\bf k}))$ as
\begin{equation}
-\alpha \Psi +2\beta_{12}|\Psi|^2\Psi - K_1D_iD_i\Psi -
2K_1C(n_iD_i)^2\Psi = 0\ .  \label{GLv}
\end{equation}
This equation has a form of the one-component GL equation for a
crystal with anisotropy axis $\bf n$ and magnetic field $\bf H$
directed perpendicular to $\bf n$. Using scaling transformation of
[$^{31}$]:
\begin{equation}
r_\parallel = r'_\parallel\,\sqrt[4]{1+2C}\ , \qquad r_\perp =
\frac{r'_\perp}{\sqrt[4]{1+2C}}\ ,
\end{equation}
we reduce (\ref{GLv}) to the isotropic GL equation
\begin{equation}
-\alpha\Psi + 2\beta_{12}|\Psi|^2\Psi -
K_1\sqrt{1+2C}\,D'_iD'_i\Psi = 0\ .      \label{GLev}
\end{equation}
 From (\ref{GLev}) we conclude the following:

\noindent
(i) the one-quantum vortex $(n=1)$ is the most favorable one;

\noindent
(ii) the sizes of the vortex corresponding to the anisotropic London
penetration length are:
\begin{equation}
\lambda_\parallel= \lambda_v\,\sqrt[4]{1+2C}\ , \qquad
\lambda_\perp=\frac{\lambda_v}{\sqrt[4]{1+2C}}\ ,
\end{equation}
where
\begin{equation}
\lambda_v = \left(\frac{\hbar^2c^2\beta_{12}}{16\pi e^2\alpha
K_1\sqrt{1+2C}}\right)^{1/2};
\end{equation}
(iii) the coherence length corresponding to Eq.\ (\ref{GLev}) is
defined as
\begin{equation}
\xi_v = \left(\frac{K_1\sqrt{1+2C}}{\alpha}\right)^{1/2};
\end{equation}
(iv) at distances larger than $\tilde\xi_v$ the vortex has the
form:
\begin{equation}
\hat{\Delta}({\bf k},{\bf r})=\Psi({\bf r})
({\bf n}\cdot\hat{\bf\Phi}({\bf k})) = \Psi(r') e^{-i\varphi'}\,({\bf
n} \cdot \hat{\bf\Phi}({\bf k}))\ ,
\end{equation}
where function $\Psi(r')$ has the same dependence as the usual Abrikosov
vortex (\ref{oneqv}). As in the case of the vortices in the broken
time-reversal phase, the characteristic scale of $\tilde \xi_v$ is
calculated by comparing the contribution to the free energy due to
admixing of the $\hat{\bf\Phi}_\pm$ component and the characteristic
gradient energy on the scale $\tilde\xi_v$. Since the density of the
gradient energy is anisotropic, the region where all components of the
order parameter are mixed is defined by the anisotropic scales
\begin{equation}
\tilde\xi_\parallel = \tilde\xi_v\,\sqrt[4]{1+2C}\ , \qquad
\tilde\xi_\perp=\frac{\tilde\xi_v}{\sqrt[4]{1+2C}}\ ,
\end{equation}
where
\begin{equation}
\tilde\xi_v =\xi_v \left(-\frac{\beta_1}{\beta_2}\right)^{1/2}.
\end{equation}

     The lower critical field for the vortices in the vector phase is
calculated from (\ref{Hc1}) by substitution $\lambda\rightarrow
\lambda_v$, $\xi\rightarrow \tilde\xi_v$, $\epsilon\rightarrow
\tilde\epsilon_v$. Note again that the above description of the long
range vortex structure is valid only if $\lambda_v \gg\tilde \xi_v$.

\subsubsection{Vortex core structure}

     The calculation of the vortex core structure for $\beta_2<0$ has
not been done yet. Therefore we will outline only expected
qualitative features of them.

     Similarly to the case of $\beta_2>0,$ we start from the
description of the ``most symmetric vortex.'' In order to find its
``full symmetry,'' we note that the rotational symmetry of the vortex
is broken due to the vector $\bf n$: $\{\hat{L}_\pi, RU_{2x}, P \}$.
The general structure of the ``most symmetric vortex'' inside the
$\tilde\xi_v$ region is given by
\begin{equation}
\hat{\Delta}({\bf k},{\bf r})=({\bf n}\cdot\hat{\bf\Phi}({\bf k}))
\psi({\bf r}) + i({\bf m}\cdot\hat{\bf\Phi}({\bf k}))\chi({\bf r})\ ,
\end{equation}
where $\bf m$ denotes the unit vector which is perpendicular to $\bf
n$ and $\bf H$. Functions $\psi({\bf r})$, $\chi({\bf r})$ have $D_2$
symmetry, and their phases vary by $2\pi$ about ${\bf r}=0$. Outside
the core region $\psi({\bf r}) \rightarrow \Psi(r')e^{-i\varphi'}$,
$\chi({\bf r}) \rightarrow 0$. Both amplitudes $\psi({\bf r})$ and
$\chi({\bf r})$ vanish at the vortex center.

     If $\tilde\xi_\parallel$, $\tilde\xi_\perp$ are of the order of
$\xi_\parallel$, $\xi_\perp$, the ``most symmetric vortex'' is the most
favorable. When $\tilde \xi_v$ becomes sufficiently large in
comparison with $\xi_v$, one can expect the appearance of vortices with
a lower symmetry as is the case with the vortices in the broken
time-reversal phase. However, this question requires further
investigations.

\subsection{Discussion}

     Summarizing this Section, let us discuss what conclusions we can
draw from the outlined picture of vortices in unconventional
superconductors and what problems they pose for further studies.

     The appearance of the unconventional vortices was mostly studied
in the framework of the two-component model when ${\bf H}\parallel{\bf
c}$. It was shown that in some range of parameters the vortices
originating from the broken time-reversal phase $(\beta_2>0)$ have the
nonaxially symmetric structure of the core, unlike the vortices in
usual superconductors which are always axisymmetric. It was also shown
that the domain structure of the ground state of a superconductor
plays an important role in its magnetic properties.

     The instability of the vortex core with axisymmetry breaking, the
domain formation and the vortex interaction with domain boundaries are
certainly the features connected to each other. However, the most
favorable ground state of such a system in a finite magnetic field
near $H_{c1}$ and, in particular, the exact symmetry of the lattice of
nonaxial vortices are still unknown.

     The vortices in a vector phase ($\beta_2<0$) were studied much
less. It is known that the vortices have an anisotropic shape in the
direction of $\hat{\bf\Phi}({\bf k})$. However, the structure of the
vortex core (depending on the parameter $\beta_2$) needs further
investigations.

     In the foregoing the vortices have been treated in the limit when a
vortex can be clearly divided into two principal regions: the {\it
region of the screening currents}, in which the main vortex energy is
concentrated, and the {\it vortex core region}, in which the
axisymmetry instabilities occur. This limit corresponds to the
restriction $\lambda \gg\tilde \xi $ which is the case whenever
$\beta_2$ is not too close to zero. Note, however, that this case is
in a sense the most trivial one because all the results were obtained
by the straightforward generalization of two famous theories:
Abrikosov theory of the vortex structure (for the region
$r>\tilde\xi$) and the theory of the vortex core instability in the
superfluid $^3$He [$^{60}$] (adopted for the two-component order
parameter $\hat{\bf\Phi}_1\pm i\hat{\bf\Phi}_2$ or $\hat{\bf\Phi}$).

     A very interesting problem comes out from the above discussion:
what is the structure of the vortices at small $\beta_2$ when the
condition $\lambda >\tilde \xi$ is not satisfied? In this case not
only the currents which originate from the gradient of phase of the
order parameter but also the currents due to the variation of all
other admixed components of the order parameter contribute to the
screening of magnetic field. Another related question is how the
vortices in the broken time-reversal phase transform to those in the
real (vector) phase when the parameter $\beta_2$ changes from
$\beta_2>0$ to $\beta_2<0$. Note that these problems require a careful
consideration because even the ground state of a two-component
superconductor is nonuniform at small $\beta_2$ ($H=0$) and has a
space-modulated laminar structure [$^{54,81}$]. Generating of
vortices in this state (and therefore the problem of $H_{c1})$ is also
not clear enough [$^{29,56}$].

     As we have seen, magnetic vortices in a two-component
superconductor have a lot of unusual features. However, the
two-component model is the simplest one which describes the
unconventional superconductivity. Therefore, one can expect even more
peculiar new features in other  multicomponent superconductors. In
this connection, it is worthwhile to mention the mixed state in the
triplet superconductor with a weak spin-orbital coupling which was
studied by Burlachkov and Kopnin [$^{12}$]. They showed that in
magnetic field spin-vectors of such a superconductor form a distinct
texture which allows the applied field to penetrate inside a
superconductor without forming singularities. It is remarkable that
the lower critical field for such a texture can be much smaller than
corresponding field for the usual singular vortices.

    \section{The Upper Critical Field in Unconventional \protect\\
    Superconductors}

     The calculation of the upper critical field for the
superconductors with multicomponent order parameter is analogous to
the case of $s$-wave superconductors (see Sec.\ 1.2). Starting from
the multicomponent GL energy functional, we derive the corresponding
linearized GL equations under assumption that magnetic field inside a
superconductor is uniform and equals to the external applied field.
Therefore, the calculation of the upper critical field reduces to the
search for eigenvalues of the system of the linearized GL equations
and then to the choice of the critical (eigen) field with the maximal
value.

     In Sec.\ 3.1 we present characteristic examples for which linearized
GL equations for multicomponent order parameter admit analytical
solutions. We will see that, despite the complicated structure, their
eigensolutions can be classified by definite quantum numbers. The
nature of this classification is a consequence of the symmetry of the
normal metal in magnetic field (Sec.\ 3.2). Hence, properties of the
superconducting phases can be discussed without solving the linearized
GL equations and general conditions of a kink on the $H_{c2}(T)$ curve
(the problem related to the phase diagram of UPt$_3$) can be
formulated (Sec.\ 3.3). In Sec.\ 3.4 we will enumerate the specific
features of the $H_{c2}$ anisotropy which make it possible to
distinguish the unconventional superconductivity from the usual one.
Finally, in Sec.\ 3.5 we consider the possibility of the appearance of
superconducting phases modulated along the field direction.

   \subsection{Upper Critical Field in the Two-Component Model}

     In this Subsection we consider two cases with ${\bf H}
\parallel{\bf c}$ and ${\bf H}\perp{\bf c}$ when linearized GL
equations for the two-component order parameter can be solved analytically
[$^{11,67,74,79}$]. The question of $H_{c2}$ anisotropy for
arbitrary directed field will be considered in Sec.\ 3.4.

     To obtain the linearized GL equation, we neglect in (\ref{GLeq2})
the forth order terms. For ${\bf H}\parallel \hat{\bf z}$ these
equations are
\begin{equation}
\alpha\!\!\pmatrix{\eta_+ \cr \eta_-}\!=\! -\! K_1\!\!
\pmatrix{(1+C)(D_x^2+D_y^2)-\frac{\displaystyle 2e}
{\displaystyle \hbar c}HD & C(D_x+ iD_y)^2 \cr
C(D_x- iD_y)^2 & (1+C)(D_x^2+D_y^2)+
\frac{\displaystyle 2e}{\displaystyle \hbar c}HD}
\!\!\pmatrix{\eta_+ \cr \eta_-},
\label{lin1}
\end{equation}
where $\eta_\pm = \eta_x \pm i \eta_y$. In (\ref{lin1}) we set
$D_z\bg{\eta}=0$, since the variation of the order parameter along
$\bf H$ only decreases $H_{c2}$. Defining the creation and
annihilation operators
\begin{equation}
a=\left(\frac{\hbar c}{4|e|H}\right)^{1/2} (D_x-iD_y)\ ,\qquad \left[a,
a^+\right]=1\ , \label{crea}
\end{equation}
one obtains the eigenfunctions of (\ref{lin1}) as combination of
the $\bf r$-dependent Landau levels wave functions $f_n({\bf r})$ (for
which $a^+a\,f_n=(n+\frac{1}{2})f_n$) and $\bf k$-dependent basis
functions $\hat{\Phi}_i({\bf k})$:
\begin{eqnarray}
\hat{\Delta}_n({\bf k},{\bf r}) & = & f_n({\bf r})
(\hat{\Phi}_x({\bf k})-i \hat{\Phi}_y({\bf k}))
+ f_{n-2}({\bf r}) (\hat{\Phi}_x({\bf k}) + i\hat{\Phi}_y({\bf k}))\ ,
\nonumber \\[2mm]
 & &\lambda_n=(2n-1)(1+C)\pm \sqrt{(2+2C-D)^2+4C^2n(n-1)}\ . \label{set}
\end{eqnarray}
Their respective eigenfields $H_n$ are related to the eigenvalues
$\lambda_n$ of the linear problem by the expression
\begin{equation}
H_n=\frac{\hbar c \alpha}{2|e| K_1\lambda_n}\ ,
\end{equation}
and $H_{c2}=\max\{H_n\}$. The solution with the lowest eigenvalue is either
\begin{eqnarray}
\hat{\Delta}_a({\bf k},{\bf r}) & = &
f_0({\bf r}) (\hat{\Phi}_x({\bf k})-i \hat{\Phi}_y({\bf k}))\ ,
\nonumber   \\
 & &\lambda_a=\lambda_0=1+C-D\ , \qquad\qquad\qquad\quad\ \,
{\rm for}\ D>\frac{C^2}{1+C}\ , \label{axial}
\end{eqnarray}
or
\begin{eqnarray}
\hat{\Delta}_{SK}({\bf k},{\bf r}) & = & f_0({\bf r}) (\hat{\Phi}_x
({\bf k}) + i \hat{\Phi}_y({\bf k})) + \omega f_2({\bf r})
(\hat{\Phi}_x({\bf k}) - i\hat{\Phi}_y({\bf k}))\ , \nonumber \\[1.5mm]
 & &\lambda_{SK} = \lambda_2=3(1+C)-\sqrt{8C^2+(2+2C-D)^2}\ ,
\nonumber \\
 & & \omega=\frac{\lambda_2-1-C-D}{2\sqrt{2}C}\ , \qquad\qquad\qquad
\quad {\rm for}\ D < \frac{C^2}{1+C}\ .   \label{SK}
\end{eqnarray}

     The new feature of the anisotropic pairing which can be seeing from
(\ref{set}) is the nonfactorized dependence of $\hat{\Delta}$ on
$\bf k$ and $\bf r$. Scharnberg and Klemm (SK) were the first who
discovered this fact solving Gor'kov equations for $p$-wave pairing
in the weak-coupling limit [$^{62}$]. These solutions were obtained
within the phenomenological Ginzburg-Landau approach in
[$^{67,79}$]. Note that the solutions (\ref{set}) contain those
products of $\bf k$- and $\bf r$-dependent functions, for which the
sum of the projection of the Cooper pair angular momentum $m$ and
Landau level number $n$ is the same. The corresponding quantum number
$N=n+m$ was introduced in [$^{37}$]. We shall call $N$ as
``generalized Landau level number.''  The symmetry reasons for the
appearance of such quantum number will be given in the next
Subsection.

     For different values of phenomenological constants $C$, $D$, the
smallest eigenvalue is either $\lambda_a$ or $\lambda_{SK}$. If the
approximate particle-hole symmetry exists near the Fermi surface
(e.g.\ in the weak-coupling regime) then $D\approx 0$. Therefore it is
the SK-phase (\ref{SK}) which should appear at $H_{c2}$ in the ``real''
two-component superconductor (see Fig.\ 4).

     Due to the cylindrical symmetry of the GL functional (\ref{GL2}),
$H_{c2}$ is isotropic in the basal plane. Taking ${\bf H}\parallel
\hat{\bf x}$, we obtain from (\ref{GLeq2}) the following system of
equations ($K=K_4/K_1$):
\begin{equation}
\alpha \pmatrix{\eta_x \cr \eta_y}
=-K_1\pmatrix{(1+2C)D_x^2+D_y^2+KD_z^2  &  2CD_xD_y  \cr
  2CD_xD_y      & D_x^2+(1+2C)D_y^2+KD_z^2}
\pmatrix{\eta_x \cr \eta_y}.                       \label{lin2}
\end{equation}
This system can be solved analytically only under the condition
$D_x\bg{\eta}=0$ (see Sec.\ 3.5) when two equations in (\ref{lin2})
become decoupled and coincide with the one-component GL equations for
anisotropic $s$-wave superconductors [$^{31}$]. Using the scaling
transformation, one obtains in this case the eigenfunctions of (\ref{lin2})
with the lowest eigenvalues [$^{11}$]:
\begin{eqnarray}
\Delta_1({\bf k},{\bf r}) & = &
f_0(s_1 y, s_1^{-1} z)\:\hat{\Phi}_x({\bf k})\ , \qquad
\lambda_1=\sqrt{K}\ ,  \label{etax}     \\
 & & s_1=\sqrt[4]{K}\ ,  \nonumber     \\
\Delta_2({\bf k},{\bf r}) & = & f_0(s_2 y,
s_2^{-1} z)\:\hat{\Phi}_y({\bf k})\ , \qquad
\lambda_2=\sqrt{K(1+2C)}\ ,     \label{etay}  \\
  & & s_2=\sqrt[4]{K/(1+2C)}\ .  \nonumber
\end{eqnarray}
Therefore, for $C>0,$ the phase (\ref{etax}) corresponds to the upper
critical field, whereas for $C<0$ the phase (\ref{etay}) appears in
the vicinity of $H_{c2}$.

     Scaled zero Landau level functions (\ref{etax}) and (\ref{etay})
are superpositions of all even Landau levels. Therefore the above
classification by quantum number $N$ reduces to $N$ over ${\rm
mod}\,2$. In addition, solutions (\ref{set}), (\ref{etax}) and
(\ref{etay}) are characterized by the parity with respect to the
reflection in the plane perpendicular to the filed direction.

    \subsection{Symmetry Classification of Superconducting Phases
near $H_{c2}$}

     The appearance of the Abrikosov vortex lattice in type-II
superconductors at $H_{c2}(T)$ is a second order phase transition.
Hence, it can be considered in the framework of the Landau phase
transition theory [$^{34}$] irrespectively of particular model
assumptions, such as the GL approximation. The corresponding  symmetry
approach  was  developed recently in the works [$^{37,40,82}$]
(see also [$^{18}$]).

     The appearance of the vortex lattice results in spontaneous
breaking of the symmetry group of the normal state in a magnetic
field. Let us first consider the Cooper pairing in an isotropic metal,
when $G=SO_3\times P$. In the presence of the magnetic field $\bf H$
directed along $\hat{\bf z}$ axis the total symmetry group $\cal G$ is
reduced to
\begin{equation}
{\cal G} = T \times D_{\infty h}(C_{\infty h}) \times
U(1)\ . \label{Gm}
\end{equation}
The group $D_{\infty h}(C_{\infty h})$ contains the subgroup
$C_{\infty h}$ of arbitrary rotations about $\bf H$ with the
reflection in the perpendicular plane $ \hat{ \sigma}_h$ and the
combined symmetry elements $RU_2$ and $R\hat{\sigma}_v$. Here $U_2$
are rotations through angle $\pi$ about axes perpendicular to $\bf H$,
and $\hat{\sigma}_v$ are reflections in planes containing $\bf H$.

     The phase transition of a continuous type implies that the order
parameter obeys the linear equation on the transition line:
\begin{equation}
\hat{\cal L}\{H,T\}\:\Delta ({\bf k},{\bf r}) = 0\ . \label{symb}
\end{equation}
This equation can be considered as a symbolic form of either
differential GL equations, e.g.\ (\ref{lin1}) and (\ref{lin2}), or
integral Gor'kov equations, see [$^{38,62}$]. The linear operator
$\hat{\cal{L}}$ is invariant under the action of the symmetry group
$\cal G$. In general, the analytic solution of (\ref{symb}) is a
complicated problem. But it is known that different solutions of
linear equations belong to different irreducible representations of
the corresponding symmetry group. This classification is model
independent and valid on the whole line $H_{c2}(T)$.

     Note that the group (\ref{Gm}) has the same structure as group
(\ref{G}) except for the breaking of time reversal symmetry. For
instance, subgroup $T$ is an abelian group of three-dimensional
translations. However, in magnetic field symmetry operators acting on
the order parameter include gauge transformations (along with the
changes of space variables), which return the chosen vector potential
$\bf A(r)$ to its original form:
\begin{eqnarray}
\hat{T}_{\bf a} &=& \exp\left\{-i\,\frac{2e}{\hbar c}\int_{\bf 0}^{\bf r}
\left[A_k({\bf r}' + {\bf a}) - A_k({\bf r}') \right]\,dr'_k
\right\}\cdot\exp({\bf a}\bg{\nabla})\ ,    \nonumber \\
\hat{L}_{\varphi} &=& \exp\left\{-i\,\frac{2e}{\hbar c}\int_{\bf 0}^{\bf r}
\left[s_{kp}A_p(s_{qj}r'_q)-A_k( r'_j)\right]\,dr'_k\right\}\cdot
\exp(i\varphi(\hat{l}_z+\hat{l}_k))\ ,          \label{mgnopr}
\end{eqnarray}
where
$$
\hat{l}_z = -i(x\partial_y-y\partial_x)\ , \qquad
\hat{l}_k = i\left(\frac{\partial}{\partial{\bf k}}\times {\bf k}
\right)\!\mbox{}_z\ , \qquad
s_{kp} = \pmatrix{\ \ \cos\varphi & \sin\varphi  \cr
-\sin\varphi & \cos\varphi}.
$$
We wrote the gauge transformation for the Cooper pair as that for a
single particle with a total charge of $2e$ only for the sake of
brevity. The exact transformation operators acting on the order
parameter (\ref{OP}) include exponents of the sum of two integrals,
both in the above form, with upper limits ${\bf r}_1$ and ${\bf r}_2$
and multiplied by charge $e$. Our simplification does not change
commutation rules for $\hat{T}_{\bf a}$. Therefore, symmetry
classification given below is valid not only in the GL regime, in
which $\bf r$ varies over distances $\xi\gg\xi_0$, but also at
$T\rightarrow 0$ when the space variations are of the order the
Cooper pair size.

     For operators (\ref{mgnopr}) the following relations hold:
\begin{equation}
\hat{T}_{\bf a}\hat{T}_{\bf b}=\hat{T}_{{\bf a}+{\bf b}}
\exp\left\{-i\,\frac{2e}{\hbar c}\int_{\bf 0}^{\bf a}
\left[A_k({\bf r}' + {\bf b}) - A_k({\bf r}')\right]\,dr'_k\right\},
\label{prod}
\end{equation}
\begin{equation}
\hat{T}_{\bf a}\hat{T}_{\bf b}=\hat{T}_{\bf b}\hat{T}_{\bf a}
\exp\left\{i\,\frac{2e}{\hbar c}\,{\bf H}\,
[{\bf a}\!\times\!{\bf b}]\right\},     \label{rel}
\end{equation}
\begin{equation}
\hat{L}_{\varphi_1}\hat{L}_{\varphi_2} =
\hat{L}_{\varphi_1+\varphi_2}\ .
\end{equation}
Although the two-dimensional translations in $x$--$y$ plane commute
with each other, this is not true for the corresponding magnetic
operators (\ref{rel}). As follows from (\ref{prod}), the operators
$\hat{T}_{\bf a}$ form not an ordinary vector representation of the
abelian group $T$, but a ray one (e.g., see [$^{13}$]). Different
choices of the gauge for the same distribution of the magnetic field
correspond to the equivalent ray representations. Near $H_{c2}$ the
superconducting order parameter is characterized by its transformation
properties under the action of magnetic operators or, equivalently, by the
corresponding irreducible ray representation.

     In order to classify the different irreducible ray
representations, we introduce the generators of infinitesimal ray
transformations:
\begin{eqnarray}
\hat{t}_x &=& -i\partial_x-\frac{2e}{\hbar c}\,(A_x+Hy)\ , \nonumber\\
\hat{t}_y &=& -i\partial_y-\frac{2e}{\hbar c}\,(A_y-Hx)\ ,
\label{genr}    \\
\hat{L}_z &=& \hat{l}_z+\hat{l}_k-\frac{2e}{\hbar c} \left(xA_y-yA_x-
\frac{H}{2}\,(x^2+y^2)\right),          \nonumber
\end{eqnarray}
which satisfy the following commutation relations:
\begin{equation}
\left[\hat{t}_x,\hat{t}_y\right] = -\frac{2e}{\hbar c}Hi\ ,\qquad
\left[\hat{t}_x,\hat{L}_z\right]=-i\hat{t}_y\ ,\qquad
\left[\hat{t}_y,\hat{L}_z\right]=i\hat{t}_x\ . \label{comrel}
\end{equation}
In the case of axially symmetric gauge ${\bf A} =
(-\frac{1}{2} Hy, \frac{1}{2} Hx,0)$ magnetic translations are reduced to
the exponent functions from usual magnetic generators [$^{9,78}$]:
\begin{equation}
\hat{t}_x = -i\partial_x-\frac{eH}{\hbar c}\,y\ ,\qquad
\hat{t}_y = -i\partial_y+\frac{eH}{\hbar c}\,x\ ,
\end{equation}
while $\hat{L}_\varphi$ is the ordinary rotation.

     By virtue of (\ref{comrel}), we introduce the Casimir operator
$\hat{\cal J}$, which commutes with operators (\ref{mgnopr}) and
operator $\hat{\cal L}$. Its eigenvalues enumerate irreducible
representations for a given ray representation:
\begin{equation}
\hat{\cal J} = \frac{\hbar c}{4 |e| H}\,(\hat{t}_x^2+\hat{t}_y^2)+
\hat{L}_z-\frac{1}{2}\ .                   \label{Cas}
\end{equation}
One can find the proof of the commutation relations of $\hat{\cal J}$
and derivation of (\ref {genr}) in [$^{82}$]. Using lowering and
raising operators (\ref{crea}), we can write $\hat{\cal J}$ in the
another form:
\begin{equation}
\hat{\cal J} = \hat{a}^+\hat{a}+\hat{l}_k\ .   \label{Casm}
\end{equation}
It is useful to construct the general type of eigenfunctions of
operator $\hat{\cal J}$ (and therefore $\hat{\cal L}$):
\begin{equation}
\hat{\Delta}_N^{\pm}({\bf k},{\bf r}) = \sum_{N=n+m} A_n(H,T)
f_n({\bf r})\hat{\Psi}_m^{\pm}({\bf k})\ ,
\label{eigenf}
\end{equation}
where $\hat{\Psi}_m^\pm({\bf k})$ is the wave function corresponding to
the definite projection $m$ of the internal angular momentum on
$\hat{\bf z}$. From the symmetry point of view, each product $f_n({\bf
r}) \hat{\Psi}_m^{\pm} ({\bf k})$ with $n+m=N$ can be considered as
one from the equivalent choices for the basis functions of
the $N$th irreducible ray representation. Then constants $A_n(H,T)$
depend on the construction of the operator $\hat{\cal L}$ in the
particular model. In addition to $N$, we define another quantum number
$\sigma=\pm$, which is the parity under reflection $\hat{\sigma}_h$ in
the plane perpendicular to $\bf H$.

     It can be easily seen that phases (\ref{set}) present a specific
case of (\ref{eigenf}). The quantum numbers $(N,\sigma)$, which were
introduced in the previous Subsection, are conditioned by the
operators $\hat{\cal J}$, $\hat{\sigma}_h$. Note that the number $N$
generalizes the Landau level number $n$: when the wave function does
not depend on the second variable ${\bf k}$, the $N$-level
classification reduces to usual Landau levels.

     In the case of anisotropic crystal, the rotational symmetry is
broken. If, however, magnetic field is directed along a $p$-fold axis,
one can see that $N$-level classification reduces to classification in
$N$ over ${\rm mod}\,p$. If magnetic field is directed in the basal
plane of hexagonal crystal (along $C_2$ axis) the above classification
is reduced to the parity of $N$, as we have seen for the
eigenfunctions (\ref{etax}) and (\ref{etay}).

     In the absence of spin-orbital coupling spin projection on the
field direction provides another quantum number, which distinguishes
irreducible representations. If the spin-orbital interaction is
strong, the mixing of singlet and triplet superconductivity is
possible in a magnetic field. From the symmetry point of view, the
space inversion $P$ transforms basis functions of a particular Landau
level through each other. Therefore, functions (\ref{eigenf}) do
not possess any parity under $P$. Hence at $H_{c2}$ the triplet states
on which the same irreducible ray representation of the group $\cal G$
is realized should arise simultaneously with the singlet states,
although the pairing type occurred at $H=0$ has near $H_{c2}$ the
dominant amplitude.

     In conclusion, one can apply the above classification scheme for
all systems described by the symmetry group (\ref{Gm}).
Two-dimensional electron systems under a magnetic field are the most
interesting among others. They were considered, e.g., in the works
[$^{32,77}$], where particular level operators analogous to Casimir
operator (\ref{Cas}) were introduced. It is general symmetry
properties of the system which make it possible such procedure in all cases:
for a single electron, interacting electrons, electrons with
spin-orbital coupling, or electron-phonon interaction.

\subsection{Kink in the Upper Critical Field}

     The peculiar $H$--$T$ phase diagram of superconducting UPt$_3$
with two jumps in specific heat at temperatures $T_{c1}$ and $T_{c2}$
and with the intersection of two critical fields $H_1(T)$ and $H_2(T)$
at the kink point (see Fig.\ 1) is the main argument in favor of the
anisotropic pairing in heavy fermions and of the multicomponent
superconducting order parameter in UPt$_3$ in particular [$^{68}$].
The splitting of phase transition is believed to be explained by the
closeness of critical temperatures of two superconducting states. The
critical fields $H_1(T)$ and $H_2(T)$ are interpreted as upper
critical fields of these phases.

     The explanation of the phase diagram of UPt$_3$ has two different
aspects: (i) what is the nature of the splitting of the phase
transition, and (ii) under what conditions the double superconducting
phase transition leads to the phase diagram similar to that in Fig.\
1.

     A comprehensive discussion of the first point can be found in
[$^{27,43,61}$]. We only enumerate the most popular models
of $T_c$ splitting in UPt$_3$.
\smallskip

\noindent (1) Antiferromagnetic (AFM) symmetry breaking field
models. (a) In some of these models [$^{24,28,45,61}$] it is
assumed that the unconventional superconductivity in UPt$_3$ is
described by the two-component order parameter
$\bg{\eta}=(\eta_1,\eta_2)$ in accordance with (\ref{expan})
which corresponds to a 2D irreducible
representation of the point crystal group $D_6$, either $E_1$ or
$E_2$. A weak AFM order with staggered moments in the basal plane
lowers the symmetry group of the system from $D_6$ to $D_2$ and
therefore splits $T_c$. (b) A more complicated variant [$^{53}$]
assumes a one-dimensional irreducible representation of $D_6$ for
superconducting order parameter but with additional degrees of freedom
due to the triplet pairing and weak spin-orbit coupling. The splitting
is due to the interaction of AFM moments with the total spin of the
Cooper pairs.
\smallskip

\noindent (2) The model of nearly isotropic $d$-wave pairing. In this
picture [$^{83}$] UPt$_3$ is a $d$-wave isotropic superconductor
with small splitting of critical temperature according to the
irreducible representations of $D_{6h}$ group. Two of these
representations $A_{1g}$ and $E_{1g}$ are responsible for the phase
transitions in UPt$_3$.
\smallskip

\noindent (3) Accidental degeneracy models [$^{19,28}$] unlike
the other models do not explain the physical reasons for $T_c$
splitting. They only suggest that the closeness of critical
temperatures of two irreducible representations of $D_{6h}$ group is
accidental.

     The answer on the second question can be given using the quantum
numbers technique of Sec.\ 3.2. The upper critical field corresponds
to the maximal eigenfield of the linearized equations. The crossing of
two eigenfields $H_1(T)$ and $H_2(T)$ leads to the kink in
$H_{c2}(T)={\rm max} \{H_1(T),H_2(T)\}$. Using the general quantum
mechanical rule that only symmetry different eigenlevels of the linear
problem can cross, we came to the following conclusion:
\begin{quote}
\it
The kink is possible only if the eigenfunctions related to $H_1(T)$
and $H_2(T)$ have different quantum numbers $N$ or $\sigma$.
Otherwise the kink should be absent or at least smeared.
\end{quote}
Application of this rule to the cited above models of $T_c$ splitting
allows to verify easily their relevance to the phase diagram of
UPt$_3$. In fact several models [$^{19,53,61,83}$]
satisfactory explain stability of the
kink in $H_{c2}(T)$ for different orientations of $\bf H$, and a new
experimental information is needed to select one of them.

\subsection{Anisotropy of the Upper Critical Field}

     In conventional anisotropic superconductors the upper critical
field can be calculated by taking into account the anisotropy of
gradient terms in the GL functional. One should replace the gradient
term in (\ref{GL1}) by $K_{ij} D^*_i \Psi^* D_j\Psi$ (in the BCS
theory, $K_{ij}$ is proportional to the inverse tensor of the
effective masses of conduction electrons). The second rank tensor
$K_{ij}$ has three independent components ($K_{xx},K_{yy},K_{zz}$) for
biaxial crystals, two components ($K_{\parallel}=K_{zz}$,
$K_{\perp}=K_{xx}=K_{yy}$) for uniaxial crystals and only one
component ($K_{xx}=K_{yy}=K_{zz}$) for cubic crystals. Therefore, in
uniaxial crystals (e.g., with hexagonal or tetragonal structure) the
angular dependence of $H_{c2}(\vartheta)$ has the form of an
axisymmetric ellipsoid with the main axes: $H^{\parallel}_{c2}=\hbar c
\alpha /2|e| K_{\perp}$, $H^{\perp}_{c2}= \hbar c \alpha /2|e|
(K_{\perp}K_{\parallel})^{1/2}$ (see Fig.\ 5). For a cubic crystal the
upper critical field in the GL region near $T_c$ is isotropic.

     Beyond the GL region the higher gradient terms slightly break the
ellipsoidal form of the angular dependence of $H_{c2}$ and restore the
anisotropy of $H_{c2}$ which coincides with the full point symmetry
group of the crystal. However, this contribution to $H_{c2}$ is of the
order of $(1-T/T_{c})^{2}$, and it does not change the slope of the
upper critical field $H'_{c2}= (dH_{c2}/dT)|_{T=T_c}$.
Recently different types of upper critical field anisotropy in
the higher orders of $(1-T/T_c)$ have been discussed for conventional
and unconventional superconductors in [$^{50}$].

     Turning back to the unconventional superconductivity, one should
expect that anisotropy properties of $H'_{c2}$ are different from
those for the $s$-wave superconductor due to a more complicated
structure of the GL equations. The problem of $H'_{c2}$ anisotropy for
different multicomponent order parameters for the crystal structures
relevant to the heavy fermions was investigated in
[$^{21,11,44}$]. The following unusual for $s$-wave
superconductors features were found:
\begin{itemize}
\item {\it For the cubic crystals} like UBe$_{13}$ (point group
$O_{h}$), $H'_{c2}$ should always have the cubic anisotropy for the
different field orientations.

\item {\it For the tetragonal crystals} like CeCu$_2$Si$_2$,
URu$_2$Si$_2$ ($D_{4h}$), the square-like anisotropy of $H'_{c2}$ is
expected when the magnetic field lies in the basal plane (see Fig.\ 6)

\item {\it For the hexagonal crystals} like UPt$_3$, UPd$_2$Al$_3$,
UNi$_2$Al$_3$ ($D_{6h}$), there are no additional features of
$H'_{c2}$ anisotropy with respect to $s$-wave superconductors.
$H'_{c2}$ is isotropic in the basal plane and has a uniaxial outplane
anisotropy.
\end{itemize}
Note that these types of ``unusual'' anisotropy are connected to the
symmetry of corresponding GL functionals. However, such anisotropy was
found for neither UBe$_{13}$ [$^{3}$], nor for CeCu$_2$Si$_2$
[$^{66}$].
Moreover, the upper critical field in CeCu$_2$Si$_2$ shows unexpected
isotropy for all field directions in the linear regime near $T_c$
[$^{66}$]. This discrepancy
can be explained by the weak sensitivity of the electron-electron
pairing interaction to the crystal anisotropy (e.g., it is the case
for the $p$-wave pairing [$^{38}$]).

     Another characteristic feature of the anisotropy of the upper
critical field in the multicomponent superconductors is the
nonelliptical, nonmonotonous behavior of $H_{c2}(\vartheta)$ for
uniaxial crystals when the applied field changes its direction from
the principal axis to the basal plane. This behavior is connected with
the existence of several solutions with close critical fields for
arbitrary directions of $\bf H$ rather than with the symmetry of GL
equations [$^{80}$]. The investigation of the GL equations for
two-component model (\ref{GLeq2}) in [$^{5,80}$] shows that the
angular dependence of $H'_{c2}$ can have pronounced minimum in a
particular region of the phenomenological parameters $C$ and $K$ (see
Fig.\ 7). Such behavior of $H_{c2}(\vartheta)$ was also found beyond
GL approximation in [$^{58}$], what confirms the universality of
this feature.

     Unfortunately, the available experimental data on the
anisotropy of the upper critical field in
UPt$_3$ [$^{69}$] are not clear enough to verify such a possibility. The
observation of the nonmonotonous behavior of $H_{c2}(\vartheta)$ at
uniaxial crystals CeCu$_2$Si$_2$, UPt$_3$, URu$_2$Si$_2$,
UPd$_2$Al$_3$, or UNi$_2$Al$_3$ would be a good argument in favor of
unusual superconductivity in heavy fermions.

     The striking feature of the superconductivity in UPt$_3$ is the
crossover in the anisotropy ratio $H_{c2}^\perp/H_{c2}^\parallel$ with
temperature, which is less than one near $T_c$ and greater than one at
$T=0$. Possible explanation of this experimental fact in terms of
unconventional superconductivity is suggested in [$^{15}$].

     At the end of this Subsection, we shall consider the quantum
number classification for solution of the linearized anisotropic GL
equations. When magnetic field is directed at an arbitrary angle with
respect to the principal axis of a uniaxial system, the point symmetry
group $D_{\infty h}(C_{\infty h})$ reduces to $G'=\{E,P,RU_{2x}\}$,
where $\hat{\bf x}$ axis lies in the basal plane perpendicularly to $\bf H$.
At the first sight, there are no quantum numbers for solutions
of linearized equations in this case because rotational symmetry about
$\bf H$ is completely broken.

     However, the GL functional constructed for a single irreducible
representation with a particular parity possesses an additional hidden
symmetry. The action of space inversion $P$ on a solution
transitionally invariant along $\bf H$ reduces to the rotation by
angle $\pi$ in the plane perpendicular to the magnetic field.
Therefore, these solutions can be described by the parity of the
Landau level number $N$, as in the case of $\bf H$ directed along the
two-fold axis. For this reason, the vortex lattice in conventional
superconductors has the form of the distorted triangular lattice with
a two-fold symmetry at all intermediate directions of $\bf H$.

\subsection{Modulation of Order Parameter Along Magnetic Field}

     The appearance of superconducting phases modulated along
field direction near $H_{c2}$ was discussed from the beginning
of investigations of upper
critical field in nontrivial superconductors [$^{11}$]. The
discussion has been renewed last time in connection with the phase
diagram of UPt$_3$ [$^{18,39}$]. Here we formulate
analytical approach to this problem on the example of two-component
model.

     If the order parameter is allowed to vary along $\bf H$, an
additional quantum number, momentum $p_H$, appears. The reflection
$\hat{\sigma}_h$ transforms functions with $p_H$ and $-p_H$ into each
other, hence, they have the same eigenvalue. In Sec.\ 3.1 we have seen
that for $\bf H\perp c$ Eq.\ (\ref{lin2}) could not be solved
analytically for arbitrary $p_x \ne 0$. Therefore, in this case we
must verify the possibility that solutions with nonzero $p_x$ have
smaller eigenvalues than (\ref{etax}) or (\ref{etay}).

     It is obvious that solutions with $p_x\gg1/l_H$ have larger
eigenvalues because magnetic field may be considered as a perturbation
to positively definite gradient terms  (\ref{cond}) in this case.
We can check analytically also an opposite
case of $p_x\ll 1/l_H$ when instead nonzero $p_x$ can be treated as a
perturbation to eigenvalues (\ref{etax}) and (\ref{etay}).

     Let us consider Hamiltonian (\ref{lin2}) as a sum of three parts
$\hat{H}_0+\hat{H}_1+\hat{H}_2$, where $\hat{H}_0$ does not depend on
$p_x$, $\hat{H}_1$ depends linearly and $\hat{H}_2$ depends
quadratically on $p_x$. Note that $\hat{H}_1$ is an off-diagonal
operator, therefore its contribution is of the order of $p_x^2$ and
negative for the lowest eigenvalue. Eigenvalues and eigenfunctions of
$\hat{H}_0$ are known:
\begin{eqnarray}
L_n & = & \frac{2|e| H K_1}{\hbar c}\sqrt{K}\,(2n+1)\ ,
\qquad\qquad\qquad           \bg{\eta}=(f_n,0)\ , \\
M_n & = & \frac{2|e|HK_1}{\hbar c}\sqrt{K(1+2C)}\,(2n+1)\ , \qquad\
\bg{\eta}=(0,g_n)\ ,
\end{eqnarray}
where $f_n$ and $g_n$ are introduced in (\ref{etax}) and (\ref{etay})
wave functions of scaled Landau levels. They are characterized by the
parity under reflection $\hat{\sigma}_x$ and by the parity under
rotation by angle $\pi$ about $\hat{\bf x}$. The variation of
solutions along $\hat{\bf x}$ breaks the classification according to
the parity under reflection, but the parity under rotation by angle
$\pi$ is still conserved. Then if we take $\eta_x$ as a superposition
of even Landau levels, $\eta_y$ must be a sum of wave functions from
odd Landau levels and vice versa.

     In accordance with this symmetry classification, the second order
perturbation theory gives the following corrections to the eigenvalues
$L_0$ and $M_0$:
\begin{eqnarray}
L_0(p_x) & = & L_0 + K_{123}\,p_x^2 - K_{23}^2\,p_x^2 \sum_{n=2k+1}
\frac{|\langle f_0| D_y |g_n\rangle|^2}{M_n-L_0}\ , \label{L0} \\[1.5mm]
M_0(p_x) & = & M_0 + K_1\,p_x^2 - K_{23}^2\,p_x^2 \sum_{n=2k+1}
\frac{|\langle g_0| D_y |f_n\rangle|^2}{L_n-M_0}\ . \label{M0}
\end{eqnarray}
Substituting $L_n$ and $M_n$ to (\ref{L0}) and (\ref{M0}), we find:
\begin{eqnarray}
L_0(p_x) - L_0 & = &  K_1(1+C)p_x^2 \left(1+\varepsilon-
2\varepsilon^2\,\sqrt{\frac{1-\varepsilon}{1+\varepsilon}}\,(1+q)^3
\sum_{k=0}^{\infty} q^{2k}\,\frac{(2k)!}{2^{2k}(k!)^2} \times \mbox{}
\right.\nonumber \\
 & & \left. \mbox{} \times \left[\frac{1}{2(1+\varepsilon)} -
\frac{\varepsilon}{((1+\varepsilon)(4k+3) - \sqrt{1-\varepsilon^2})
(1+\varepsilon + \sqrt{1-\varepsilon^2})} \right] \right),
\nonumber \\[2mm]
 & & \varepsilon=\frac{C}{1+C}\ , \qquad
q=\frac{\varepsilon}{1+\sqrt{1-\varepsilon^2}}\ ,
\end{eqnarray}
and the same expression for $M_0(p_x)-M_0$ with the only change
$\varepsilon \rightarrow -\varepsilon$. If $C>0$, then
$L_0$ is the smallest eigenvalue for solutions uniform in
the field direction. A lower estimate on $L_0(p_x)$ is obtained
neglecting the second term in square brackets. Then, using
$\sum_{k=0}^\infty q^{2k}\frac{(2k)!}{2^{2k}(k!)^2} = \frac{1}
{\sqrt{1-q^2}}$, we come to:
\begin{equation}
L_0(p_x) - L_0 > K_1(1+C)p_x^2 \left(1+\varepsilon-
\varepsilon^2\sqrt{\frac{1-\varepsilon}{(1+\varepsilon)^3}}\,
\frac{(1+q)^3}{\sqrt{1-q^2}} \right). \label{compar}
\end{equation}
It may be easily checked that the right hand side of Eq.\
(\ref{compar}) is always positive for any $C>0$ ($0<\varepsilon<1$).
The same is true for function $M(p_x)$ in the opposite case when
$-0.5<C<0$ ($-1<\varepsilon<0$).
Thus, we may conclude that modulation of superconducting phase near $H_{c2}$
along $\bf H$ are unfavorable in both cases $p_x\ll 1/l_H$ and $p_x\gg 1/l_H$.
It is very improbable that such modulations will appear for
intermediate
values of $p_x$.

     While a solution with the lowest eigenvalue, e.g.\ $L_0$ for
$C>0$, is  stable against small nonhomogeneous perturbations with $p_x\ne
0$, this is not always true for the next level $M_0(p_x)$. It follows
from (\ref{M0}) that when $M_0 \rightarrow L_1$ (i.e., $C \rightarrow
4$) the corresponding denominator tends to zero, giving a large
negative contribution to the coefficient at $p_x^2$. Thus, at $C=4$
the level $M_0$ is certainly unstable against modulations with
particular wavelength. The numerical investigation of $M_0(p_x)$ shows
that coefficient at $p_x^2$ changes sign from $+$ to $-$ at $C=0.946$.

     A well developed modulated structure at $C>0.946$ can not be
analyzed on the basis of the second order perturbation theory. An
infinite system of secular equations must be solved in this case.
Variational calculations show that for $C=4$ ($\varepsilon=4/5$) when
$M_0=3$ (in of units $2|e|HK_1\sqrt{K}/\hbar c$) the next level after
$L_0=1$ has the eigenvalue $M_0(p_x^*)=2.24$ with the modulation
wavelength  $\lambda=2\pi/p_x^* \sim 4.7\,l_H$. The discussed instability
of the second solution of the linearized GL equations is important
for consideration of phase transitions inside mixed state (see Sec.\ 5).

 \section{Vortex Lattices Near $H_{c2}$}

The next necessary step in the investigation of the superconducting phase
transition at the upper critical field is determination
of the vortex lattice form near $H_{c2}$. We describe in this Section
nontrivial features of this procedure applied to the multi-component
solutions of linear equations like SK-phase (\ref{SK}).

     The symmetry approach to second order phase transitions
[$^{34}$] allows us to make a few general conclusions before
fulfilling calculations. First of all, the free energy near $H_{c2}$
has the form of the Landau expansion in powers of $\hat{\Delta}$.
The structure of the order parameter $\hat{\Delta}({\bf k},{\bf
r})$ is determined by the solution of the linear equations (\ref{symb}) and
corresponds to the particular irreducible ray representation $(N,
\sigma)$. The superconducting amplitude $\hat{\Delta}$ is
proportional to $(H_{c2}-H)^{1/2}$. The degeneracy of the order
parameter is lifted by the forth order terms of the Landau functional.
While the second order term is unique and can be always presented
as $|\hat{\Delta}_N^\pm|^2$, the forth order terms, generally, become
nonlocal at low temperatures.
For this reason, we consider only a restricted problem ---
investigation of nonlinear equations in GL approximation.

     We derive below the energy parameter for multicomponent
GL functionals which generalies corresponding
Abrikosov formula (Sec.\ 4.1). In
Sec.\ 4.2 we discuss general procedure of finding the most stable
lattice form and consider symmetry properties of the energy parameter
under different choices of the lattice parametrasation. Sec.\ 4.3
is devoted to the search of the energetically favorable forms of
the Abrikosov lattice for the SK-phase ({\ref{SK}), which
should appear near $H_{c2}$ in the two-component
superconductor. Symmetry group
of the vortex lattice is constructed in Sec.\ 4.4.

\subsection{Average Energy Density}

     In this Subsection we construct the energy parameter for an
unconventional superconductor, which determines the form of the vortex
lattice. For this purpose we must take the order parameter in an
unchanged form as a solution of the linearized GL equations with the
largest critical field. The relative accuracy of this approximation is
$(1-H/H_{c2})$. The vector potential ${\bf A}_0$ (${\rm rot}\,{\bf
A}_0={\bf H}_{c2}$), which enters the linearized GL equations, must be
changed in order to include the diamagnetic field of superconducting
currents: ${\bf h}={\rm rot}\,{\bf A= H+h}_s$, ${\rm rot}\,{\bf h}_s =
4\pi/c \cdot {\bf j}_s$. Neglecting by the boundary effects, we come to
the following functional:
\begin{equation}
F = \frac{H-H_{c2}}{4\pi}\,\langle|h_s|\rangle +
\langle|\hat{\Delta}|^4\rangle - \frac{H^2}{8\pi} -
\frac{\langle h_s^2\rangle}{8\pi}\ ,                    \label{L1}
\end{equation}
where $\langle\ldots\rangle$ denotes spatial average. The forth order
terms in (\ref{L1}) are written in a symbolical form, which implies
the sum over all the possible invariants constructed from the
components $\eta_i$.

     Thus the calculation of $F$ reduces to the derivation of the
diamagnetic field $h_s$ with the help of the Maxwell equations
[$^{16}$]. For one-component superconductors, in which spatial
dependence of the order parameter is restricted to zero Landau level
functions, this field may be found analytically: $h_s= 8\pi e
K|\Psi({\bf r})|^2 / \hbar c$.

     Since the calculation of $h_s$ for the multicomponent SK-phase
becomes more difficult, we restrict ourselves to the case of
$\kappa\gg1$. This allows us to neglect by the last term in
(\ref{L1}), which is always of the order of $\langle |\hat{\Delta}
|^4 \rangle / \kappa^2$. Thus we should find only
$\langle|h_s|\rangle$. In the chosen approximation $h_s$ is the
quadratic form of $\hat{\Delta}_N({\bf k},{\bf r})$. The magnetic
moment of the unit volume is a quantity invariant under action of the
symmetry group (\ref{Gm}). Therefore, the average $\langle |h_s|
\rangle$ is the second order invariant constructed from
$\hat{\Delta}({\bf k},{\bf r})$, and, consequently,
$\langle|h_s|\rangle \propto \langle|\hat{\Delta}_N({\bf k},{\bf
r})|^2\rangle$. The numerical factor in this relation does not depend
on the particular structure of solution, because formally at
$H=0$ the first term in (\ref{L1}) must be equal to $-\alpha\langle|
\hat{\Delta} |^2 \rangle$. Thus for large $\kappa$ the
multicomponent GL energy is:
\begin{equation}
F = \mbox{} - \alpha\left(1-\frac{H}{H_{c2}}\right)
\langle|\hat{\Delta}|^2\rangle +
\langle|\hat{\Delta}|^4\rangle - \frac{H^2}{8\pi}\ .
\label{L2}
\end{equation}
Minimizing (\ref{L2}) by the amplitude of $\hat{\Delta}$, we get
\begin{equation}
F = -\frac{\alpha}{4\delta} \left(1-\frac{H}{H_{c2}}\right)^2
 - \frac{H^2}{8\pi}\ ,
\end{equation}
where
\begin{equation}
\delta = \frac{\langle|\hat{\Delta}|^4\rangle}
{\langle|\hat{\Delta}|^2\rangle^2}
\label{enpar}
\end{equation}
is the energetic parameter of GL model which generalies
usual Abrikosov parameter for ordinary superconductors.
The absolute minimum of
$\delta$ determines the geometrical form of the vortex lattice near
$H_{c2}$. For conventional superconductors it is
\begin{equation}
\delta_0 = \beta\,\frac{\langle|f_0|^4\rangle}
{\langle|f_0|^2\rangle^2}\ ,                      \label{par1}
\end{equation}
in accordance with the Abrikosov formula [$^{1}$]. Vortex lattice form
for axial phase
(\ref{axial}) of the two-component model is determined by the same parameter
(\ref{par1}) with $\beta$ changed on $\beta_1$. Substituting (\ref{SK}) in
(\ref{enpar}) we obtain that for the SK-phase the
energy parameter has the following form:
\begin{equation}
\delta_{SK} = \frac{\langle\beta_1(|f_0|^4+\omega^4|f_2|^4)+2
\omega^2(\beta_1 + 2\beta_2)|f_0|^2|f_2|^2\rangle}
{(1+\omega^2)^2\langle|f_0|^2\rangle^2}\ .               \label{par2}
\end{equation}
We discuss energy minimums of these parameters in the Sec.\ 4.3.

       \subsection{Properties of Lattice Solutions}

Before search for the favorable lattice form we construct in this
Section general periodic solutions on different Landau
levels which then should be substituted in (\ref{enpar})--(\ref{par2}).
We will choose particular parametrisation for two-dimensional
vortex lattices and discuss its symmetry properties which are
helpful in further numerical calculations.

     The vortex lattice is defined by the discrete group of
translations through $\bf a$ and $\bf b$:
\begin{equation}
\hat{T}_{\bf a}\:\hat{\Delta}({\bf k},{\bf r}) = \exp(i\varphi_1)\:
\hat{\Delta}({\bf k},{\bf r})\ , \qquad
\hat{T}_{\bf b}\:\hat{\Delta}({\bf k},{\bf r}) = \exp(i\varphi_2)\:
\hat{\Delta}({\bf k},{\bf r})\ , \label{trcon}
\end{equation}
with arbitrary phase factors $\varphi_1$ and $\varphi_2$. It follows
from (\ref{trcon}) that magnetic translations for this subgroup
commute. Thus, due to (\ref{rel}) any periodical solution possesses an
integer number of flux quanta per unit cell [$^{9,78}$].

     We consider only lattices with one quantum of flux per unit cell.
All distances are measured  in the units of the magnetic length
$l_H^2=\hbar c/2|e|H$, and the vector potential is chosen in the
Landau gauge ${\bf A}=(-Hy,0,0)$. Using the explicit form of the
magnetic operator in this gauge $\hat{T}_{\bf a} = \exp(-ia_y
x)\exp({\bf a}\bg{\nabla})$, it is possible to construct the lattice
solution on the $n\/$th Landau level
\normalsize
\begin{eqnarray}
f_n({\bf r}\,|\,\tau) & = & \sum_m\:\exp\left\{-\pi i\rho m^2+\pi
im(\rho+1)+\frac{2\pi i}{a}\left(m-\frac{1}{2}\right)x -
\frac{1}{2}\left[y-\left(m-\frac{1}{2}\right)b\sin\alpha\right]^2
\right\}\times \nonumber \\
 & & {}\times H_n\left[y-\left(m-\frac{1}{2}\right)b\sin\alpha\right],
\label{period}
\end{eqnarray}
which satisfies (\ref{trcon}) with ${\bf a}=(a,0)$, ${\bf b} = (b
\cos\alpha, b \sin\alpha)$, $\varphi_1=\pi$, $\varphi_2=\pi(\rho+1)$.
The area of the unit cell is fixed by the condition of flux
quantization: $a b\sin\alpha=2\pi$. The lattice form depends on
the complex parameter $\tau = \rho +i\sigma = b/a\,\exp(i\alpha)$. The
parameters $\tau_h=\exp(\pi i/3)$ and $\tau_s=i$ correspond to
lattices with hexagonal and square symmetries. The periodic function
$f_0({\bf r}\,|\,\tau)$ has one zero per unit cell located at $(0,0)$.

     The functions (\ref{period}) can be written as:
\begin{equation}
f_0({\bf r}\,|\,\tau) = \overline{\vartheta_1
\left(\left.\frac{z}{a}\,\right|\tau\right)}\,
\exp\left(-\frac{y^2}{2}\right), \qquad
f_n({\bf r}\,|\,\tau)\propto(a^+)^n f_0({\bf r}\,|\,\tau)\ ,
\end{equation}
where $z=x+iy$, $\vartheta_1(z\,|\,\tau)$ is the Jacobi
theta-function, and the bar denotes complex conjugation. The change of
the vector potential affects only the exponential factor in the
expression for $f_0({\bf r}\,|\,\tau)$, leaving the $\tau$-dependence
through $\overline{\vartheta_1(z/a\,|\,\tau)}$ for all gauges. The
theta-functions are the only quasiperiodical analytical functions.
This leads to the uniqueness of $f_n({\bf r}\,|\,\tau)$ with given
periods $\bf a$, $\bf b$ and phases $\varphi_1$, $\varphi_2$.
Substituting (\ref{period}) into (\ref{eigenf}), we obtain all
possible one-quantum vortex lattices near $H_{c2}$.

     In order to find which type of lattice is realized, one should
substitute $\hat{\Delta}_N({\bf k},{\bf r}\,|\,\tau)$ into the
nonlinear energy functional, calculate the free energy $F(\tau)$ for
each $\tau$, and select $\tau$ which corresponds to the absolute
minimum of $F(\tau)$. Before calculations we establish some symmetry
properties of function $F(\tau)$ according to [$^{82}$], which
generalizes ideas of [$^{59}$].

     The choice of the basis for two-dimensional Brav\'e lattice is
not unique. The same Brav\'e lattice may be described by different
$\tau$. Instead of $\bf a$ and $\bf b$ any pair of their integer
combinations $k{\bf a} + l{\bf b}$ and $m{\bf a} + n{\bf b}$, such
that $kn-lm=\pm 1$, can be taken as basis vectors. This transformation
changes parameter $\tau$ to $\tau' = (m + n\tau)/(k + l\tau)$. The
matrices $\pmatrix{k & l \cr m & n }$, which do not change the
orientation of the basis form the modular group $SL(2,{\rm Z})$.
Together with substitution ${\bf a}\rightarrow -{\bf a}$, ${\bf
b}\rightarrow {\bf b}$ they represent a complete set of possible
parameterizations of the original Brav\'e lattice.

    In a physical problem, the  symmetry of the periodical system can
be lower than the symmetry of the corresponding Brav\'e lattice. The
state is specified not only by the lattice, but also by additional
parameters describing the structure and orientation of the each
lattice site. For such systems transformation in the space of its
parameters that consists only of a different choice of basis vectors
changes the initial state. Thus we are dealing with the presence or
absence of the symmetry under the action of the group $SL(2,{\rm
Z})\times {\rm Z}_2$ in the complex plane of parameter $\tau$.

     Functions $f_n({\bf r}\,|\,\tau)$ (or $\hat{\Delta}_N ({\bf
k},{\bf r}\,|\,\tau)$) are uniquely determined by their lattice, and,
consequently, they should be symmetrical under $SL(2,{\rm Z})\times
{\rm Z}_2$ group. This statement can be verified by establishing the
following properties of $\hat{\Delta}_N({\bf k},{\bf r}\,|\,\tau)$
as a function of $\tau$ (see proof in [$^{83}$]):
\begin{equation}
\hat{\Delta}_N ({\bf k},{\bf r}\,|\,\tau + 1) =
\hat{\Delta}_N ({\bf k},{\bf r}\,|\,\tau)\ ,    \label{prop1}
\end{equation}
\begin{equation}
\hat{\Delta}_N ({\bf k},{\bf r}\,|\,-1/\tau) = \sqrt{\tau^*}\,
e^{-iN\varphi}\, \exp\left[\frac{\pi i}{4}
\left(\rho-3-\frac{\rho}{\rho^2+\sigma^2}\right)\right]
\hat{L}_\varphi\:\hat{\Delta}_N ({\bf k},{\bf r}\,|\,\tau)\ ,
\label{prop2}
\end{equation}
where $\exp(i\varphi)=\sqrt{\tau/\tau^*}$, and
\begin{equation}
\hat{\Delta}_N ({\bf k},{\bf r}\,|\,-\tau^*) = RU_{2y}\
\hat{\Delta}_N ({\bf k},{\bf r}\,|\,\tau)\ .   \label{prop3}
\end{equation}

     Substituting $\hat{\Delta}_N ({\bf k},{\bf r}\,|\,\tau)$ to
an arbitrary functional $F$ invariant under group $\cal G$, we obtain
energy $F(\tau)$, which is not changed under $SL(2,{\rm Z})\times{\rm
Z}_2$ transformations:
\begin{equation}
F(\tau) = F\left( \frac{m+n\tau}{k+l\tau}\right) = F(-\tau^*)\ .
\end{equation}
The fundamental region of $SL(2,{\rm Z})\times{\rm Z}_2$ group in
complex plane contains all different one-quantum lattices. A specific
example of the fundamental region is shown in Fig.\ 8. Their vertices
$\tau_h$ and $\tau_s$ are always stationary points of $F(\tau)$. While
at the point $\tau_s$ functional can achieve all types of extremums, in
$\tau_h$ only a local maximum or a local minimum are possible. The
absolute minimum of $F(\tau)$ can be located, of course, not only at
these symmetrical points, but somewhere in the fundamental region.
Moreover, such exotic behavior is characteristic for unconventional
pairing (see Sec.\ 4.3).

     It is interesting that similar theoretical constructions have
appeared independently in the theory of Weinberg-Salam model for the
electroweak interactions [$^{41,42}$]. Vacuum reconstruction in
high magnetic fields in this theory is described by energy functional
on complex wave function $\Psi$ with nonlocal forth order term arising
from coupling with additional bosonic field. This nonlocality in the
forth order terms could results in its turn in exotic forms of lattice
solutions.

     \subsection{Vortex Lattice in Two-Component Model}

     The explicit dependence of the energy parameter for one-component
superconductors from $\tau$ can be obtained after
substitution of $f_0({\bf r}\,|\,\tau)$ into (\ref{par1}). This
calculation was done in [$^{59}$]:
\begin{equation}
\delta_0(\rho,\sigma) = \beta\sqrt{\sigma}\sum_{m,n}
\cos(2\pi\rho m n)\: e^{-\pi\sigma(m^2+n^2)}\ .           \label{d0}
\end{equation}
Here summation is fulfilled over all integer $m$ and $n$. The absolute
minimum of the function $\delta_0(\rho,\sigma)$ is achieved at the
point $\tau_h$ of the fundamental region which corresponds to the
regular triangular lattice.

     The energy parameter of the SK-phase (\ref{par2}) can be presented
as:
\begin{equation}
\delta_{SK}(\rho,\sigma) =
\frac{\beta_1}{(1+\omega^2)^2}\, \left(\delta_{00}(\rho,\sigma) +
2 \omega^2(1+2\beta) \delta_{02}(\rho,\sigma) +
\omega^4 \delta_{22}(\rho,\sigma)\right),
\label{d1}
\end{equation}
where parameter $\omega$ is given by (\ref{SK}) and $\beta=\beta_2/\beta_1$.
Explicit expressions for $\delta_{pq}(\rho,\sigma)$ can be found
in Appendix. Positions of the absolute minimums are
different for each function $\delta_{pq}(\rho,\sigma)$. Therefore,
variation of phenomenological parameters $\beta$ and $\omega$ leads to
the rich phase diagram for the vortex lattice form near $H_{c2}$. Note
that, according to (\ref{SK}), the following restriction
$0<\omega<1/\sqrt{2}$ is fulfilled in the region of the existence of
SK-phase near $H_{c2}$.

    The numerical investigation of $\delta_{SK}(\rho,\sigma)$ was done
first in [$^{48}$]. But the phase diagram found there
contradicts in fact general requirements of the phenomenological theory,
because it predicts two coinciding phase transitions lines between
triangular and square lattices and between square and rectangular
lattices. The correct
phase diagram for the position of the absolute minimum of function
$\delta_{SK}(\rho,\sigma)$ versus phenomenological parameters $\beta$
and $\omega^2$ is presented in Fig.\ 9. There are different regions
where the energetically favorable solution corresponds to the regular
triangular, square, rectangular, and distorted triangular lattices.
Phase transitions lines in Fig.\ 9 between square and the
rectangular lattices and between square and distorted triangular
lattice IV are of the second order, all other transitions are of the
first order. For negative $\beta$ the only stable form of the vortex
lattice is a triangular one.

    The characteristic scale of the difference between various
extremums of $\delta_{SK}(\rho,\sigma)$ is of the order of
$10^{-4}$--$10^{-6}$ compared to $10^{-2}$ for
$\delta_0(\rho,\sigma)$. Note that the point
$(\beta,\omega^2)=(0.5,0.1)$ predicted by BCS theory for $E_{1g}$
representation is situated in the region of stability of the hexagonal
lattice. It is interesting that for $\delta_{SK}(\rho,\sigma)$ the
absolute minimum always lies on the boundary of the fundamental
region.

    In Sec.\ 4.2 we have argued that the one-quantum vortex lattice
with hexagonal symmetry always corresponds to the extremum of the
energy functional for arbitrary superconducting order parameter
$\hat{\Delta}_N({\bf k},{\bf r})$. Now we can see from Fig.\ 9 that
corresponding stability region of the hexagonal lattice
for the SK-phase is rather large.

      \subsection{Symmetry Group of the Vortex Lattice}

To complete the discussion of vortex lattices near $H_{c2}$
we should write explicitly corresponding symmetry groups,
which include magnetic translations and rotations (\ref{mgnopr})
in combinations with nontrivial phase factors.

     The transformation rules of $\hat{\Delta}_N({\bf k},{\bf r})$
near $H_{c2}$ under magnetic translations by the lattice periods are
defined by (\ref{trcon}) and (\ref{period}). Vortex lattices
corresponding to different phases $\varphi_1$ and $\varphi_2$ go over
into each other under magnetic translations which do not coincide with
any integer combination of the basis vectors. Displaced solutions
$\hat{T}_{\bg{\rho}} \hat{\Delta}_N({\bf k},{\bf r}\,|\,\tau)$ are
transformed like Bloch wave functions with wave vectors lying in the
Brillouin zone of the background lattice $\hat{\Delta}_N({\bf
k},{\bf r}\,|\,\tau)$ [$^{40}$].

     We are interested mainly in the transformation properties
of the triangular lattice, since it has a particular region of stability
for arbitrary $N$. Using identity $-1/\tau_h=\tau_h$ one can obtain
from (\ref{prop1})--(\ref{prop3}) following properties under
rotation by angle $\pi/3$ about arbitrary
vortex axis $\Gamma$:
\begin{equation}
\hat{L}_{\pi/3}\:\hat{\Delta}_{N,\Gamma}({\bf k},{\bf r}\,|\,\tau_h)
= e^{-i\pi(1-N)/3} \hat{\Delta}_{N,\Gamma}({\bf k},{\bf
r}\,|\,\tau_h)\ ,
\end{equation}
and under time-reversal together with rotation by angle $\pi$ about
an axis perpendicular to the direction of magnetic field:
\begin{equation}
RU_{2y}\:\hat{\Delta}_{N,\Gamma}({\bf k},{\bf r}\,|\,\tau_h) =
 \hat{\Delta}_{N,\Gamma}({\bf k},{\bf r}\,|\,\tau_h)\ .
\end{equation}
Combining the elements listed above, we can write the symmetry group
of the vortex lattice as
\begin{equation}
{\cal G}_{\Gamma}=\left\{ e^{i\pi}\hat{T}_{\bf a},\ \
e^{i\pi/2}\hat{T}_{\bf b},\ \ e^{i\pi(1-N)/3}\hat{L}_{\pi/3},\ \
\sigma\hat{\sigma}_h,\ \  RU_{2y} \right\}.     \label{Gg}
\end{equation}
All other elements are formed as proper combinations of this
basis. For example, the superconducting order parameter
$\hat{\Delta}_{N,\Gamma}({\bf k},{\bf r}\,|\,\tau_h)$ also
possesses symmetries under rotations by angle $2\pi/3$ about an axis
passing through the center of the triangle $K$ and on angle $\pi$
about the middle of the triangle side $M$ (see Fig.\ 10). Considering
two other displaced vortex lattices $\hat{\Delta}_{N,K}({\bf
k},{\bf r}\,|\,\tau_h)$ and $\hat{\Delta}_{N,M}({\bf k},{\bf
r}\,|\,\tau_h)$, whose points $K$ and $M$ coincide with the center of
coordinate $\Gamma$, we can establish their symmetries with respect to
$\Gamma$:
\begin{equation}
{\cal G}_{K}=\left\{ e^{-i\pi/3}\hat{T}_{\bf a},\ \
e^{-i\pi/6}\hat{T}_{\bf b},\ \ e^{-2i\pi N/3}\hat{L}_{2\pi/3},\ \
\sigma\hat{\sigma}_h,\ \  e^{i\pi}RU_{2y} \right\},
\label{GK}
\end{equation}
\begin{equation}
{\cal G}_{M}=\left\{ e^{i\pi}\hat{T}_{\bf a},\ \
e^{-i\pi/2}\hat{T}_{\bf b},\ \ e^{i\pi N}\hat{L}_\pi,\ \
\sigma\hat{\sigma}_h,\ \  e^{i\pi}RU_{2y} \right\}.\label{GM}
\end{equation}

     The symmetry properties of square lattice
$\hat{\Delta}_{N,\Gamma}({\bf k},{\bf r}\,|\,\tau_s)$ and of all others
from Sec.\ 4.3 can be investigated by the similar way.

     We have derived transformation properties (\ref{Gg})--(\ref{GM})
using explicit form of $\hat{\Delta}_N({\bf k},{\bf r}\,|\,\tau)$
which is valid only for a constant magnetic field inside a sample,
i.e., when $H \rightarrow H_{c2}$. Below $H_{c2}$ the magnetic field
is modulated by superconducting currents. However, Eqs. (\ref{mgnopr})
allows introduction of discrete magnetic translations and rotations in
this case too. Therefore, the mixed state is invariant under
symmetry group (\ref{Gg}) in the finite region $H^*<H<H_{c2}$ until a
new phase transition occurs.

\section{Phase Transitions in the Mixed State}

Perhaps the most unusual feature of the mixed state of an unconventional
superconductor is the possibility of the phase transitions between different
vortex lattices.

As we have mentioned in Sec.\ 1.2, in usual superconductors
the character of vortex interaction
strongly depends on the distance between vortices. In the vicinity of
the lower critical field $H_{c1}$ vortex lattice is formed through
the interactions between nearest neighbors. At fields $H>H_c$ vortex
interacts with all neighbors within circle of the radius $\lambda$.
If, however, $H\ll H_{c2}$ the average distance between vortices
$r_L$ is much greater than coherence length $\xi$ and the intersection
of cores of different vortices is negligible. In this case only long range
hydrodynamics repulsion participates in the lattice formation.

As vortices in unconventional superconductors have the same structure
at distances $r>\tilde\xi$, all above arguments are valid for them too.
The cores intersection becomes important for intervortex distance
$r_L\sim {\rm max}\{\xi,\tilde\xi\}$. Therefore all deviations from the usual
Abrikosov theory of vortex lattice formation can occur only in the field
region near the upper critical field.
The convenient tool for investigation of vortex lattices in that region
is symmetry approach developed in Secs.\ 3 and 4.

In Sec.\ 5.1 we formulate analytical approach to the investigation of
phase transitions near $H_{c2}$ and show that the closeness of two
critical (eigen) fields in the linear problem always leads to additional
phase transitions inside mixed state. Sec.\ 5.2 is devoted to the symmetry
consideration of structural phase transitions in the vortex lattices.
Finally, in Sec.\ 5.3 we briefly discuss possible second order phase
transitions below upper critical field on the examples of two-component
GL functional (\ref{GL2}) and accidental degeneracy models.

\subsection{General Approach}

     The space distribution of the order parameter $\hat{\Delta}
({\bf k},{\bf r})$ at a given temperature $T$ and an external magnetic
field $H$ ($H_{c1}<H<H_{c2}$) is defined by the minimum of the energy
functional $F\{\hat{\Delta}({\bf k},{\bf r}), H, T\}$. To minimize
$F$, one needs to solve the nonlinear (GL) equations: $\delta F/\delta
\hat{\Delta}^* =0$, $\delta F/\delta {\bf A}=0$ which  can be
symbolically written as:
\begin{equation}
\hat{L}\{\hat{\Delta} ({\bf k},{\bf r}),H,T\} = 0\ .  \label{nonlin}
\end{equation}
Their solution near $H_{c2}(T)$ is the Abrikosov lattice with
a particular symmetry group ${\cal G}_\Gamma$.

     Suppose that at some critical field $H^*(T)$ the second order
structural phase transition in the vortex lattice does take place.
Speaking in terms of the phase transition theory, $H^*(T)$ is the
bifurcation line below which Eq.\ (\ref{nonlin}) has two different
solutions: the old one $\hat{\Delta}_1({\bf k},{\bf r})$ invariant
under the group ${\cal G}_\Gamma$ and the new one
$\hat{\Delta}_1({\bf k},{\bf r}) + \hat{\Delta}_2({\bf k},{\bf
r})$ with a lower symmetry $\widetilde{\cal G}$. At $H<H^*(T)$ the
energy functional changes the character of the extremum for solution
$\hat{\Delta}_1({\bf k},{\bf r})$ from the absolute minimum to the
local maximum, whereas $\hat{\Delta}_1({\bf k},{\bf r}) +
\hat{\Delta}_2({\bf k},{\bf r})$ corresponds to its minimum. The
small perturbation $\hat{\Delta}_2({\bf k},{\bf r})$ belongs to
some nonunit irreducible representation of the group ${\cal
G}_\Gamma$. Since the phase transition is of the second order, the
amplitude of $\hat{\Delta}_2({\bf k},{\bf r})$ at $H<H^*$ is small
as $(H^* - H)^{1/2}$. Linearizing (\ref{nonlin}) in the small value
$\hat{\Delta}_2$, one obtains the equation for the $H^*(T)$ as a
condition on the magnetic field at which the linear eigenvalue problem
\begin{equation}
\hat{\cal L} \{ \hat{\Delta}_1({\bf k},{\bf r}),H^*,T\} \:
\hat{\Delta}_2({\bf k},{\bf r}) = 0              \label{nlin}
\end{equation}
has a solution. The linear operator $\hat{\cal L}$ depends on the space
distribution of the order parameter $\hat{\Delta}_1({\bf k},{\bf
r})$ which in its turn is the solution of (\ref{nonlin})  taken at the
phase transition line $H^*(T)$.

     To specify the influence of the background solution
$\hat{\Delta}_1({\bf k},{\bf r})$ on the appearance of
$\hat{\Delta}_2({\bf k},{\bf r})$, we separate the part $\hat{\cal
L}_0$ from the operator $\hat{\cal L}$ which does not depend on
$\hat\Delta$ and present (\ref{nlin}) in the following form:
\begin{equation}
\hat{\cal L}_0\{H^*(T)\}\:\hat{\Delta}_2({\bf k},{\bf r}) +
\hat{\cal L}_1\{\hat{\Delta}_1({\bf k},{\bf r}), H^*(T)\} \:
\hat{\Delta}_2({\bf k},{\bf r}) = 0\ .
\label{divnlin}
\end{equation}
In the vicinity of $H_{c2}$ the second term in (\ref{divnlin}) is small
and can be considered as a perturbation to the equation
\begin{equation}
\hat{\cal L}_0\{H,T\}\: \hat{\Delta}({\bf k},{\bf r})=0\ ,
                                        \label{restrlin}
\end{equation}
which in fact coincides with (\ref{symb}). The last equation has
different critical fields $H_i(T)$, and the maximal of them $H_1(T)$
corresponds to $H_{c2}(T)$. Let us assume that there exists another
eigenfield $H_2(T)$ of Eq.\ (\ref{restrlin}) which is sufficiently
close to $H_1(T)$. Then one can derive similarly to (\ref{L2}) the
following {\it
two-order parameter functional\/}:
\begin{equation}
F = \mbox{} - \alpha_1\left(1-\frac{H}{H_1}\right)
\langle|\hat{\Delta}_1|^2\rangle +
\langle|\hat{\Delta}_1|^4\rangle -
\alpha_2\left(1-\frac{H}{H_2}\right)
\langle|\hat{\Delta}_2|^2\rangle +
\langle|\hat{\Delta}_2|^4\rangle + \langle\hat{\Delta}_1^2
\hat{\Delta}_2^2\rangle\ ,                      \label{2OP}
\end{equation}
which represents free energy for fields close to $H_{c2}$.

     Due to the infinite degeneracy of both order parameters, finding
all extremums of the energy functional (\ref{2OP}) is a rather
complicated problem even in the case when spatial dependencies of both
$\hat{\Delta}_1$ and $\hat{\Delta}_2$ are described by zero
Landau level functions. However, without interaction between
$\hat{\Delta}_1$ and $\hat{\Delta}_2$ the problem of the phase
transitions described by functional (\ref{2OP}) becomes trivial. There
will be two successive second order phase transitions at $H=H_1$ and
$H=H_2$. Turning on of the interaction term $\langle
\hat{\Delta}_1^2 \hat{\Delta}_2^2 \rangle$ affects only
the lower transition. At least in the finite range of parameters where
interaction term is small, the type
of the lower transition is not changed and the critical
field $H_2(T)$ is slightly renormalized to $H^*(T)$, which corresponds
to the structural phase transition in the vortex lattice. This case is
considered in the two following Subsections.

     If other eigenfields of (\ref{restrlin}) are far from
$H_{c2}(T)$, this analytic approach is not applicable, because
Eq.\ (\ref{divnlin})
can not be reduced to (\ref{restrlin}) due to the significant
admixture of all other Landau levels.
However, the continuous change of parameters of the system,
which move $H_2$ from $H_1$ to lower fields, raises a hope that in
some range of parameters the phase transition in the mixed state still
exists.

\subsection{Structural Phase Transitions}

     According to the Landau theory, if the second order phase
transition in the vortex lattice occurs, the initial symmetry cannot
be broken to an arbitrary subgroup of ${\cal G}_\Gamma$. The residual
symmetry group $\widetilde{\cal G}\subset {\cal G}_\Gamma$ is defined
as the symmetry of the new lattice $\hat{\Delta}_1({\bf k},{\bf
r}) + \hat{\Delta}_2({\bf k},{\bf r})$. To classify the possible
ways of symmetry breaking in the vortex lattice
$\hat{\Delta}_1({\bf k},{\bf r})$ one should (i) find all
irreducible representations of the group ${\cal G}_\Gamma$, (ii)
decompose the new order parameter $\hat{\Delta}_2({\bf k},{\bf
r})$ over the basis functions of a particular representation, (iii)
write the corresponding Landau functional for chosen representation,
and (iv) determine all its minimums with their residual symmetry
groups $\widetilde{\cal G}$.

     We do not intend to develop here this general approach. Staying
on the point of view of the above Subsection, we only show how this
symmetry analysis must be applied in the particular case of the energy
functional (\ref{2OP}).

     Previously we have shown that the appearance of the phase
transition in the vortex lattice at $H=H^*(T)$ should be connected to
some eigenlevel of (\ref{restrlin}). The quantum numbers
$(N_2,\sigma_2)$ of $\hat{\Delta}_2({\bf k},{\bf r})$ are different
from quantum numbers $(N_1,\sigma_1)$ of $\hat{\Delta}_1({\bf k},{\bf
r})$. Therefore, these solutions are orthogonal:
$\langle\hat{\Delta}_1^* \hat{\Delta}_2\rangle = 0$. Moreover,
the existence of a structural phase transition at $H=H^*(T)$ on a
nonunit irreducible representation of the group ${\cal G}_\Gamma$ also
means that $\langle|\hat{\Delta}_1 |^2\hat{\Delta}_1^*
\hat{\Delta}_2\rangle = 0$. Otherwise nuclei of
$\hat{\Delta}_2$ would arise at $H=H_{c2}(T)$ simultaneously with
the vortex lattice $\hat{\Delta}_1$, as is the case for all Landau
level functions with $n\:({\rm mod}\,6)=0$ in conventional
superconductors [$^{35}$].

     Irreducible representations of two-dimensional space group ${\cal
G}_\Gamma$ are characterized by the wave vector $\bf q$ lying in the
Brillouin zone of the vortex lattice. To obtain a set of corresponding
Bloch functions from those belonging to the Landau level $N_2$, we
should construct
one-quantum lattices $\hat{\Delta}_2$ of the same form as
$\hat{\Delta}_1$. Then, as we have discussed in Sec.\ 4.4, there
exists one-to-one correspondence between the displacement of
periodical solutions $\hat{\Delta}_2$ in the unit cell of
$\hat{\Delta}_1$ and the wave vector $\bf q$ in the Brillouin
zone.

     According to the Lifshitz criteria, only symmetrical points of
the Brillouin zone are important for the structural phase transitions
in a crystal lattice [$^{34}$]. For the regular triangular lattice
these are the center of the hexagon, its vertex, and the center of the
hexagon side. These points of the Brillouin zone correspond to the
following displacements: $\bg{\rho}_{\Gamma}=0$;
$\bg{\rho}_{K,K'}=\pm({\bf a} + {\bf b})/3$; $\bg{\rho}_{M,M',M''} =
{\bf a}/2, {\bf b}/2, ({\bf a} - {\bf b})/2$. The existence of
equivalent displacements is a consequence of rotational symmetry
around $\Gamma$. In the case of a distorted triangular lattice, e.g.\
when $\bf H \perp c$, this degeneracy is lifted.

     The shape of the stable lattice structure $\hat{\Delta}_1 +
\hat{\Delta}_2$ on the transition line $H^*(T)$ is determined by
interaction in the quartic terms: $\langle|\hat{\Delta}_1|^2
|\hat{\Delta}_2|^2\rangle$ and, if it exists,
$\langle\hat{\Delta}_1^{*2} \hat{\Delta}_2^2 +
c.\,c.\,\rangle$. Depending on the sign of these interactions, all
relative locations between $\hat{\Delta}_1$ and
$\hat{\Delta}_2$: $\Gamma$--$\Gamma$, $\Gamma$--$K$, $\Gamma$--$M$
are possible.

     The residual symmetry group of the mixed state at $H<H^*$ is
determined by
\begin{equation}
\widetilde{\cal G}= {\cal G}_{\Gamma}^{N_1} \cap{\cal G}_{Q}^{N_2}
\end{equation}
with appropriate $Q=\Gamma$, $K$ or $M$.

     If the vortices of the new lattice coincide with those of
$\hat{\Delta}_1({\bf k},{\bf r})$, only rotational symmetry is
broken below $H^*$. This implies the {\it distortion} of the initial
hexagonal lattice.

     If new vortices appear between the old ones in $M$ position, the
operator $e^{-i\varphi_2}\hat{T}_{\bf b}$ transforms
$\hat{\Delta}_1^{(\Gamma)} + \hat{\Delta}_2^{(M)}$ into
$\hat{\Delta}_1^{(\Gamma)} - \hat{\Delta}_2^{(M)}$ (see
(\ref{Gg}) and (\ref{GM})). In this case the unit cell of the whole
lattice is two times larger than separate cells of lattices
$\hat{\Delta}_1$ and $\hat{\Delta}_2$. It follows from the
same arguments that $\hat{\Delta}_1^{(\Gamma)} +
\hat{\Delta}_2^{(K)}$ carries three quanta of the magnetic flux
per unit cell. Such {\it period multiplications} can be either
complete or partial when elementary translations $\hat{T}_{\bf a}$ or
$\hat{T}_{\bf b}$ belong to $\widetilde{\cal G}$ in combination with
proper rotations or reflections.

\subsection{Examples}

\subsubsection{Closeness of two eigenfields}

     For $s$-wave superconductors the upper critical field always
corresponds to the zero Landau level. The first Landau level has a
critical field three times as small. For multicomponent superconductor
it is possible situation when eigenfields of different phases lie sufficiently
close to each other. This leads to structural
phase transitions inside the mixed state.

     To illustrate this case, we use the example of the two-component
superconducting order parameter. As we have seen in Sec.\ 3.1,
depending on the constants $C$ and $D$ two different phases can arise
near $H_{c2}$ for $\bf H \parallel c$. For example, if $\lambda_{SK} <
\lambda_a$, the SK-phase with the generalized Landau level number
$N_1=+1$ appears at $H_{c2}=H_{SK}$. If, however, $\lambda_a$ is
slightly larger than $\lambda_{SK}$, the critical field $H_a$ for the
axial phase with $N_2=-1$ is close to $H_{c2}$. When the regular
triangular lattice is favorable for SK-phase, the nuclei of axial phase do not
appear simultaneously with SK-phase at $H_{c2}$ because triangular lattices
for phases with different $N$ are invariant under different hexagonal
groups ${\cal G}^N_\Gamma$ (\ref{Gg}).

     To investigate possible phase transitions in the
vicinity of $H_{c2}$ one should use energy functional (\ref{2OP})
with $\alpha_1=\alpha_2$, nonlinear terms (\ref{par1}) and
(\ref{par2}), and the interaction term of the form ($\kappa \gg 1$):
\begin{equation}
\langle\hat{\Delta}_a^2 \hat{\Delta}_{SK}^2\rangle =
2\beta_1\omega^2\langle|f_0^a|^2|f_2^{SK}|^2\rangle + (\beta_1 +
2\beta_2) \langle|f_0^a|^2|f_0^{SK}|^2\rangle \ .  \label{td}
\end{equation}
According to the results of [$^{82}$], the instability at $H=H^*$
(minimum of (\ref{td})) corresponds to the appearance of $\Gamma$--$K$
structure if $\beta_2/\beta_1 > - 0.5 - 0.12\,\omega^2$ with the
residual symmetry group
\begin{equation}
\widetilde{\cal G} = \left\{ e^{i\pi/2}\hat{T}_{\bf a+b},\ \
e^{-i\pi/2}\hat{T}_{2{\bf a-b}},\ \ e^{-2i\pi/3}\hat{L}^{(K)}_{2\pi/3},\ \
\sigma\hat{\sigma}_h,\ \  RU_{2y} \right\}.         \label{tri}
\end{equation}
If $- 0.5 - 0.12\,\omega^2 > \beta_2/\beta_1 > -0.5-0.35\,\omega^2$ new
vortices are located at $M$ positions, and
\begin{equation}
\widetilde{\cal G} = \left\{ e^{i\pi}\hat{T}_{\bf a},\ \
e^{i\pi}\hat{T}_{2\bf b},\ \ \sigma\hat{\sigma}_h,\ \  RU_{2y}
\right\}.                \label{dva}
\end{equation}
If $-0.5-0.41\,\omega^2 > \beta_2/\beta_1 > -1$ two lattices
$\Delta_{SK}$ and $\Delta_a$ are not displaced. Their symmetry group
is
\begin{equation}
\widetilde{\cal G}=\left\{ e^{i\pi}\hat{T}_{\bf a},\ \
e^{i\pi/2}\hat{T}_{\bf b},\ \ \hat{L}_{\pi},\ \
\sigma\hat{\sigma}_h,\ \  RU_{2y} \right\}.
\end{equation}
For parameters $-0.35\,\omega^2 > \beta_2/\beta_1+0.5 >
-0.41\,omega^2$ the minimum is achieved for nonsymmetric displacements
between $M$ and $\Gamma$ points.

     One can see that the period multiplication in the first two cases
is complete. That is, e.g., the diamagnetic field of superconducting
currents $h_s$ is periodical only under translations from (\ref{tri})
and (\ref{dva}).

     The phase transition at $H=H^*$ leads to the phase
diagram of the two-component GL functional (\ref{GL2}) presented
schematically in Fig.\ 4.

     For $\bf H\perp c$ there are also two phases (\ref{etax}) and
(\ref{etay}) with close values of critical fields whenever $C$ is
small. Since these phases differ in the quantum number $\sigma$ and in
the parity of $N$, their mixing is prevented at $H_{c2}$. Possible
structural phase transitions in this case were partially considered in
[$^{26,82}$]. The complete analysis over the entire region of
phenomenological parameters must take into account the possibility of
the appearance of phases modulated along $\bf H$ (see Sec.\ 3.5), and
has not been done yet. Note that the available experimental data on
the vortex lattice restructuring at $H=H^*(T)$ in UPt$_3$ correspond
to this orientation of magnetic field relative to the crystal axes
[$^{30}$].

     The example in this Subsection shows that for any pairing type
(even in an anisotropic crystal with the magnetic field directed
along the rotational axis) in some range of coefficients of energy
functional on the $H$--$T$ phase diagram there exist a phase
transition line at $H_{c1}(T) <H^*(T) <H_{c2}(T)$. How large is this
region is the question of numerical calculations.

\subsubsection{Intersection of critical fields belonging to the
different pairing types}

     Let us assume that at $H=0$ two different superconducting states
with close critical temperatures $T_{a}$ and $T_{b}$ ($T_{a}>T_{b}$)
do occur. Phases $a$ and $b$ may belong to the different pairing types
$(s+p$ or $s+d)$ in an isotropic metal or to the two irreducible
representations of the point group $G$ in anisotropic crystal. Here we
will not discuss the physical reasons for this accidental closeness of
$T_{a}$ and $T_{b}$ but restrict ourselves to the question: what is
the $H$--$T$ phase diagram of the system in this case?

     The coexistence of two superconducting phases in magnetic field
is described by some GL energy functional:
\begin{equation}
F = F_a\{\hat{\Delta}_a\} + F_b\{\hat{\Delta}_b\} +
F_{ab}\{\hat{\Delta}_a,\hat{\Delta}_b\}\ ,    \label{split}
\end{equation}
which takes the form of (\ref{2OP}) near the upper critical field. The
solution of the problem of $H$--$T$ phase diagram is trivial if there
is no interaction $F_{ab}\{\hat{\Delta}_a, \hat{\Delta}_b\}$:
it will be the sum of two independent diagrams for the phases
$\hat{\Delta}_a$, $\hat{\Delta}_b$ with the upper critical
fields $H_a$, $H_b$, and under the condition $dH_b/dT>dH_a/dT$ it will
resemble the phase diagram of UPt$_3$ (Fig.\ 11).

     Being described by different irreducible representations of the
group $G$ at $H=0$, the phases $\hat{\Delta}_a$,
$\hat{\Delta}_b$ can also belong to the eigensolutions of linearized
equations of a different symmetry. Following Sec.\ 3.2, we should
attribute the different quantum numbers $(N_a,\sigma_a)$,
$(N_b,\sigma_b)$ to the critical fields $H_a$, $H_b$. The interaction
term $F_{ab}\{\hat{\Delta}_a,\hat{\Delta}_b\}$ does not change
the symmetry classification of the eigensolutions of the linear
problem. It leads to the renormalization of $H_a(T)$ and $H_b(T)$
curves inside the mixed state but does not smear the kink in
$H_{c2}(T)=\max \{H_a(T),H_b(T)\}$. Phase diagrams for different
combinations of irreducible representations for $\hat{\Delta}_a$
and $\hat{\Delta}_b$ based on Landau expension near $H_{c2}$
(\ref{2OP}) have been partially considered in
[$^{28,19,83}$]. Note, that without mixing of $\Delta_a$ and
$\Delta_b$ in the higher order gradient terms the model (\ref{split})
exhibit only partial breaking of translational symmetry below $H^*(T)$.

     The field intersection is also possible in the above example of
2D order parameter. The small uniaxial anisotropy in the basal plane
$(|\eta_x|^2 - |\eta_y|^2)$ splits $T_c$ into: $T_c^+$ for
$\hat\Phi_y({\bf k})$ phase and $T_c^-$ for $\hat\Phi_x({\bf k})$
phase. If the magnetic field is directed along $\hat{\bf x}$, there
exist an intersection of two critical fields: $H^y(T) \sim
(T_c^+-T)/(K_{123}K_4)^{1/2}$ and $H^x(T) \sim (T_c^--T) /
(K_1K_4)^{1/2}$. In fact it is the first model proposed for the phase
diagram of UPt$_3$ [$^{24,45}$].
\vspace{0.5mm}

     Concluding this Section, we want to emphasize that the quantum
number technique is very important for establishing the nature of the
phase transition at $H^*(T)$ as well as for investigation of the
stability of the kink. The exact sequence of the phase transitions for
arbitrary strength of the interaction term in the energy
functional (\ref{2OP}) is still
unknown. By numerical calculations in [$^{48}$] it was shown the
possibility of a
phase transition of the first order for large strength of interaction,
which occurs before the discussed second order structural transitions.
\vspace{3mm}

     To summarize, we have reviewed the equilibrium properties of the
mixed state of superconductors with the multicomponent order
parameter. The main feature of unconventional pairing is the reduced
symmetry of the superconducting state in comparison to the normal one.
This results in a large variety of unusual properties such as power
law temperature dependencies of thermodynamic characteristics,
domain structure due to the presence of degenerate states, the angular
dependence of the Josephson current, and others, which were considered
in details in previous reviews [$^{22,65}$]. But perhaps the most
pronounced breakdown of various symmetries occurs when vortices
penetrates inside the volume of a multicomponent superconductor. We
have considered nonaxisymmetric vortices and flux lattices of
different forms. Structural phase transitions in the Abrikosov lattice
are the main qualitatively new feature of unconventional
superconductivity which, in our opinion, should be primarily used for
establishing unusual superconductivity experimentally.

\section*{ACKNOWLEDGMENTS}

     We are grateful to S.\,V.\,Iordansky, K.\,Maki, J.\,A.\,Sauls,
M.\,Sigrist, K.\,Scharnberg and especially to V.\,P.\,Mineev and
G.\,E.\,Volovik for fruitful discussions on theoretical subject and to
M.\,Boukhny, N.\,van Dijk, J.\,Flouquet, L.\,Taillefer, A.\ de Visser
for stimulating discussions about experimental situation of
superconductivity in heavy fermion compounds. We thank
K.\,Ueda for careful reading of the manuscript and for many
improving remarks. This article was done in
part at the Institut f\"ur Theoretische Physik, RWTH Aachen, Germany
(by I.A.L.\ and by M.E.Zh.), at ISI, Torino, Italy (by I.A.L.) and at
the Nordic Institute for Theoretical Physics (NORDITA), Copenhagen,
Denmark (by M.E.Zh.). The work on this review was partially supported by
Grant No.\ MGI000 of International Science Foundation.

\section*{APPENDIX}

     The energy parameter (\ref{enpar}) for unconventional
superconductors is constructed in general case from the repeated
blocks which are integrals of periodic functions from different
Landau levels. In the case of the SK-phase they are [$^{80}$]:
\begin{eqnarray}
\delta_{00}(\rho,\sigma) =
\frac{{\displaystyle\langle|f_0|^4\rangle}}{{\displaystyle
\langle|f_0|^2\rangle^2}} & = &
\sqrt{\sigma}\:\sum_{m,n}\:\cos(2\pi\rho m n)\,
e^{-\pi\sigma(m^2+n^2)}\ ,\\
\delta_{02}(\rho,\sigma) =
\frac{{\displaystyle\langle|f_0|^2|f_2|^2\rangle}}
{{\displaystyle\langle|f_0|^2\rangle \langle|f_2|^2\rangle}} & = &
\frac{\sqrt{\sigma}}{2}\: \sum_{m,n}\: \cos(2\pi\rho m n)\,
e^{-\pi\sigma(m^2+n^2)}\left(\frac{3}{4}-\pi\sigma(m^2+n^2)
 \right.       \nonumber\\
& &  \left.\frac{}{} + \pi^2\sigma^2(m^2-n^2)^2\right),\\
\delta_{22}(\rho,\sigma) =
\frac{{\displaystyle\langle|f_2|^4\rangle}}
{{\displaystyle\langle|f_2|^2\rangle^2}} & = &
\frac{\sqrt{\sigma}}{4}\:
 \sum_{m,n}\:\cos(2\pi\rho m n)\,e^{-\pi\sigma(m^2+n^2)}
\left(\frac{41}{16} - \frac{13}{2}\pi\sigma(m^2+n^2)
\right. \nonumber\\
 & & \mbox{} + 3\pi^2\sigma^2(m^2+n^2)^2 +
\frac{19}{2}\pi^2\sigma^2(m^2-n^2)^2 -     \\
& & \left.\frac{}{} - 6\pi^3\sigma^3(n^2+m^2)(n^2-m^2)^2
+ \pi^4\sigma^4(n^2-m^2)^4\right). \nonumber
\end{eqnarray}
Behavior of these functions on different lines in the fundamental
region is shown in Fig.\ 12. The corresponding absolute minimums are
located at the following points: for $\delta_{00}(\rho,\sigma)$ at
$\rho = 0.5$, $\sigma = \sqrt{3}/2$, for $\delta_{02}
(\rho,\sigma)$ at $\rho = 0$, $\sigma = 2.6$, for
$\delta_{22}(\rho,\sigma)$ at $\rho = 0.5$, $\sigma =
\sqrt{3}/2$.

\newpage

\section*{List of Figures}

\begin{description}

\item[Fig.\ 1.] Phase diagram of superconducting states in UPt$_3$ for
(a) $\bf H\parallel c$ and (b) $\bf H\perp c$ (after [$^{2}$]).

\item[Fig.\ 2.] Contour plot of the order parameter modulus for
nonaxisymmetric vortices in the two-component model with the GL
parameters $\beta_2 = 0.1 \beta_1$ and $K_1=K_2=K_3$; (a) triangular
vortex for $\widehat{\Delta} \sim (\hat{\bf\Phi}_1 - i
\hat{\bf\Phi}_2) e^{-i\varphi}$, (b) ``crescent'' vortex for
$\widehat{\Delta} \sim (\hat{\bf\Phi}_1 + i \hat{\bf\Phi}_2)
e^{-i\varphi}$ (after [$^{70}$]).

\item[Fig.\ 3.] Phase diagram of axisymmetry instability at the
vortex core in the two-component model (after [$^{70}$]).

\item[Fig.\ 4.] The eigen critical fields and schematic $H$--$T$ phase
diagram for $\bf H\parallel c$ in the two-component model when $D <
C^2 / (1+C)$. For $D > C^2 / (1+C)$ the order of eigenfields is
reversed.

\item[Fig.\ 5.] Outplane anisotropy of $H_{c2}(\vartheta)$ in uniaxial
$s$-wave superconductors; $K_\parallel$ : $K_\perp$ = (a) 1 : 2, (b) 2
: 1, (c) 4 : 1.

\item[Fig.\ 6.] Anisotropy of $H_{c2}(\varphi)$ in the basal plane for
$E_1$ irreducible representation of $D_4$ group, $K_5$ is the coefficient
at the ``tetragonal'' invariant in gradient terms \cite{Burl} and
$K_2=K_3=K_1$;
$K_1$ : $K_5$ = (a) 2 : 1, (b) 1 : 1, (c) 1 : 2.

\item[Fig.\ 7.] Angular dependence of $H_{c2}(\vartheta)$ between $\bf
c$-axis and the basal plane in the two-component model (5),
$\varepsilon = C/(1+C)$, $K=K_4/K_1$; (a) $\varepsilon = 0.3$,
$K=0.8$, (b) $\varepsilon = 0.3$, $K=1$, (c) $\varepsilon = 0.3$,
$K=1.2$, (d) $\varepsilon = 0.2$, $K=1$.

\item[Fig.\ 8.] The fundamental region of the $SL(2,{\rm Z})\times {\rm
Z}_2$ group in the complex plane of the parameter $\tau = \rho + i
\sigma$.

\item[Fig.\ 9.] Phase diagram of the vortex lattice form near $H_{c2}$
for SK-phase. Region I corresponds to the regular triangular lattice,
region II --- to the square lattice, region III --- to the rectangular
lattice, regions IV and V --- to the distorted triangular lattices
with $45^\circ < \alpha < 60^\circ$ and $\alpha > 60^\circ$ ($\alpha <
30^\circ$) respectively. Point corresponds to the weak-coupling
parameters. Details of the phase boundaries intersections are
shown on (b).

\item[Fig.\ 10.] Symmetry elements of the triangular vortex lattice.

\item[Fig.\ 11.] Schematic $H$--$T$ phase diagram of the accidental
degeneracy model.

\item[Fig.\ 12.] Behavior of functions (a) $\delta_{00}(\rho,\sigma)$,
(b) $\delta_{02}(\rho,\sigma)$, (c) $\delta_{22}(\rho,\sigma)$ on
lines $\rho = 0$ and $\rho = 0.5$ in the fundamental region.

\end{description}
\end{document}